\documentclass{article}
\usepackage{arxiv}
\usepackage{hyperref}
\usepackage{latexsym}
\usepackage{url}
\usepackage[utf8]{inputenc}
\usepackage{booktabs}
\usepackage{amsmath}
\usepackage{amsfonts}
\usepackage{amssymb}
\usepackage[inline]{enumitem}
\usepackage{enumitem}
\usepackage{mathbbol}
\usepackage[T1]{fontenc}
\usepackage{mathtools}
\usepackage{multirow}
\usepackage{natbib}
\usepackage{rotating}
\usepackage{subcaption}
\usepackage{graphicx}
\usepackage{etaremune}
\usepackage{authblk}
\usepackage{tikz}

\newcommand{\ie}{\emph{i.e.}}
\newcommand{\eg}{\emph{e.g.}}

\newcommand{\vs}{\emph{vs. }}

\title{Overview of the TREC 2021 deep learning track}

\author[1]{Nick Craswell}
\author[1]{Bhaskar Mitra}
\author[2]{Emine Yilmaz}
\author[3,5]{Daniel Campos}
\author[4]{Jimmy Lin}
\affil[1]{Microsoft\\\texttt{\small \{nickcr, bmitra\}@microsoft.com}}
\affil[2]{University College London\\\texttt{\small emine.yilmaz@ucl.ac.uk}}
\affil[3]{University of Illinois Urbana-Champaign\\\texttt{\small dcampos3@illinois.edu}}
\affil[4]{University of Waterloo\\\texttt{\small jimmylin@uwaterloo.ca}}
\affil[5]{Neural Magic Inc}

\begin{document}
\maketitle

\begin{abstract}
This is the third year of the TREC Deep Learning track.
As in previous years, we leverage the MS MARCO datasets that made hundreds of thousands of human annotated training labels available for both passage and document ranking tasks.
In addition, this year we refreshed both the document and the passage collections which also led to a nearly four times increase in the document collection size and nearly $16$ times increase in the size of the passage collection.
Deep neural ranking models that employ large scale pretraininig continued to outperform traditional retrieval methods this year.
We also found that single stage retrieval can achieve good performance on both tasks although they still do not perform at par with multistage retrieval pipelines.
Finally, the increase in the collection size and the general data refresh raised some questions about completeness of NIST judgments and the quality of the training labels that were mapped to the new collections from the old ones which we discuss in this report.
\end{abstract}
\section{Introduction}
\label{sec:intro}
At TREC 2021, we hosted the third TREC Deep Learning Track continuing our focus on benchmarking ad hoc retrieval methods in the large-data regime.
As in previous years~\citep{craswell2019overview, craswell2020overview}, we leverage the MS MARCO datasets~\citep{bajaj2016ms} that made hundreds of thousands of human annotated training labels available for both passage and document ranking tasks.
In addition, this year we refreshed both the document and the passage collections which also led to a nearly four times increase in the document collection size and nearly $16$ times increase in the size of the passage collection.
In addition to evaluating retrieval methods on the larger collections, the data refresh also aimed to provide additional metadata---\eg, passage-to-document mappings---that may be useful for ranking as well as incorporate some fixes for known text encoding issues in previous versions of the datasets.

This year, in addition to focusing on TREC-style blind evaluation of neural methods against strong traditional baselines, the track also encouraged participating groups to annotate their runs based on whether they employ dense retrieval methods and whether their ranking pipeline is a single stage retrieval process.
The goal was to both encourage more explorations of neural methods in first stage retrieval as well as to allow analysis of how these emerging methods compare to previous state-of-the-art.

Deep neural ranking models that employ large scale pretraininig continued to outperform traditional retrieval methods this year.
We also found that single stage retrieval can achieve good performance on both tasks although they still do not perform at par with multistage retrieval pipelines.
Finally, the increase in the collection size and the general data refresh raised some questions about completeness of NIST judgments and the quality of the training labels that were mapped to the new collections from the old ones which we discuss later in this report.
\section{Task description}
\label{sec:task}

Similar to previous years, Deep Learning Track in 2021 has two tasks: Document retrieval and passage retrieval. Participants were allowed to submit up to three runs for each task. When submitting each run, participants indicated what external data, pretrained models and other resources were used, as well as information on what style of model was used.

Across the two tasks, same set of $477$ queries were used by the participants, who were allowed to submit up to three runs per task. From the full set of $477$ queries, stratified sampling based on query length was used to select the subset queries for pooling and judging. Queries were split into two strata based on their length, where queries containing more than or equal to 10 words were put into the stratum corresponding to long queries and the rest of the queries were put into the stratum corresponding to short queries. An equal number of queries were sampled from each stratum.

In the pooling and judging process, NIST chose a subset of these sampled queries for judging, based on budget constraints and with the goal of finding a sufficiently comprehensive set of relevance judgments to make the test collection reusable~\citep{craswell2021trec}. This led to a judged test set of $57$ queries ($28$ short, $29$ long queries) for the document retrieval task and $53$ queries ($25$ short, $28$ long queries)  for the passage retrieval task. 

Below we provide more detailed information about the document retrieval and passage retrieval tasks, as well as the datasets provided as part of these tasks. 

\subsection{Document retrieval task}

The first task focuses on document retrieval, with two subtasks:
\begin{enumerate*}[label=(\roman*)]
    \item Full retrieval and
    \item top-$100$ reranking.
\end{enumerate*}

The full retrieval subtask models the end-to-end retrieval scenario, documents can be retrieved from the full document collection provided and the runs are expected to rank documents based on their relevance to the query. 

In the reranking subtask, participants were provided with an initial ranking of $100$ documents, which was retrieved using Pyserini~\citep{lin2021pyserini}. This way the reranking subtask allows all participants to start from the same starting point and to focus on learning an effective relevance estimator, without the need for implementing an end-to-end retrieval system. It also makes the reranking runs more comparable, because they all rerank the same set of $100$ candidates.



For evaluation, judgments were collected on a four-point scale:
\begin{etaremune}[start=3]
    \item \textbf{Perfectly relevant:} Document is dedicated to the query, it is worthy of being a top result in a search engine.
    \item \textbf{Highly relevant:} The content of this document provides substantial information on the query.
    \item \textbf{Relevant:} Document provides some information relevant to the query, which may be minimal.
    \item \textbf{Irrelevant:} Document does not provide any useful information about the query.
\end{etaremune}
For metrics that binarize the judgment scale, we map document judgment levels 3,2,1 to relevant and map document judgment level 0 to irrelevant.

\subsection{Passage retrieval task}
Similar to the document retrieval task, the passage retrieval task includes
\begin{enumerate*}[label=(\roman*)]
    \item a full retrieval and
    \item a top-$100$ reranking tasks.
\end{enumerate*}

In the full retrieval subtask, given a query, the participants were expected to retrieve a ranked list of passages from the full collection based on their estimated likelihood of containing an answer to the question.
Participants could submit up to $100$ passages per query for this end-to-end retrieval task.

In the top-$100$ reranking subtask, $100$ passages per query were provided to participants, giving all participants the same starting point. Similar to the document retrieval subtask, the $100$ passages provided to the participants generated using Pyserini~\cite{lin2021pyserini}. Participants were expected to rerank the 100 passages based on their estimated likelihood of containing an answer to the query.

For evaluation, judgments were collected on a four-point scale:
\begin{etaremune}[start=3]
    \item \textbf{Perfectly relevant:} The passage is dedicated to the query and contains the exact answer.
    \item \textbf{Highly relevant:} The passage has some answer for the query, but the answer may be a bit unclear, or hidden amongst extraneous information.
    \item \textbf{Related:} The passage seems related to the query but does not answer it.
    \item \textbf{Irrelevant:} The passage has nothing to do with the query.
\end{etaremune}
For metrics that binarize the judgment scale, different than the document retrieval task, we map passage judgment levels 3,2 to relevant and map document judgment levels 1,0 to irrelevant.
\section{Datasets}
\label{sec:data}

This year we introduced the MS MARCO v2 dataset, which was used in both tasks. To understand how the new dataset differs from the old, we will first describe the natural language generation data and v1 ranking data.

\begin{figure}
    \centering
    \includegraphics[width=0.75\linewidth]{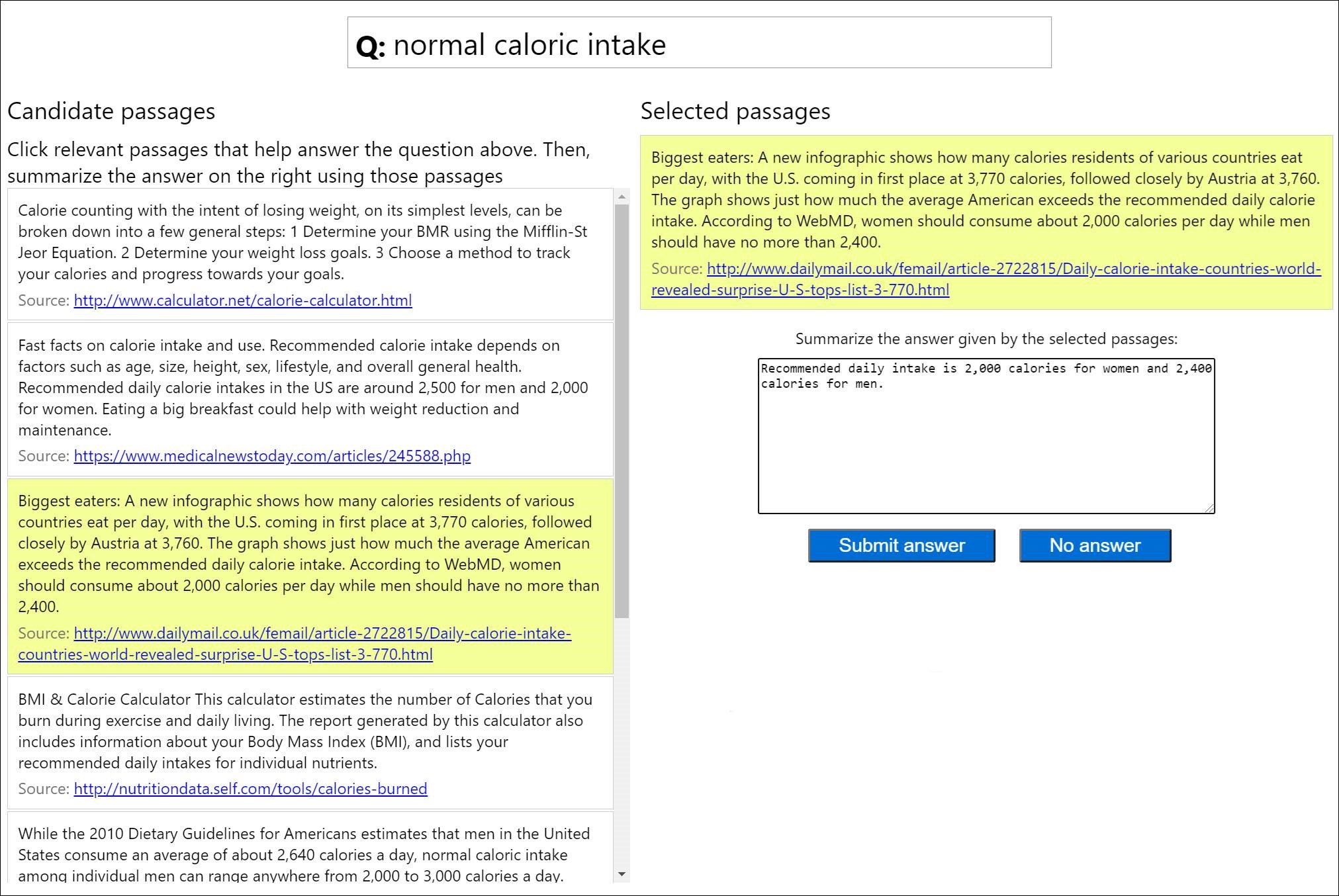}
    \caption{Crowd task used to generate the original MS MARCO natural language generation leaderboard. This same crowd data was later adapted to become the MS MARCO ranking tasks.}
    \label{fig:ms_marco_hit}
\end{figure}

\paragraph{MS MARCO natural language generation dataset.}
The original MS MARCO dataset was for a natural language generation task, rather than a ranking task. It processed one million queries, using a crowd task as shown in Figure~\ref{fig:ms_marco_hit}. The crowd worker would read the query, consider up to ten passages related to the query, decide if the passages could be used to answer the query and if answerable write an answer to the query in their own words. For each answerable question the crowd works provided a non-extractive answer and an annotation of which passages they used to generate their answer. There was substantial quality work with the crowd workers to ensure quality and the crowd workers spent an average of 2.5 minutes on each annotation. The million queries represent a real Bing query workload, and the ten results come from a Bing component designed to handle that workload. The queries were filtered before being annotated to remove any adult or offensive queries and any non-English queries. Moreover, further filtering was performed to ensure that the queries came from the $10-20\%$ of English queries that were detected as potentially being answerable with a short passage. Although the filter may be imperfect, the intention was to exclude navigational queries (such as [youtube]), queries that require a longer answer (such as [beef wellington recipe]) and queries that aim to complete some transaction (such as [buy xbox live]). We note that about 35\% of the queries could not be answered using the ten passages, in which case the crowd worker would indicate No answer, and one part of the original MS MARCO challenge was to predict which queries were answerable.

\paragraph{MS MARCO ranking v1 datasets.}
The MS MARCO passage and document ranking v1 datasets are used in the current MS MARCO leaderboards~\citep{lin2022fostering, craswell2021ms, lin2021significant} and in TREC 2019 and TREC 2020.

To generate the v1 passage ranking data, we took the union of the top ten passage lists for the one million queries, giving us 8.8 million distinct passages. For queries that were answerable, we used the crowd judge annotation for selected passages as a positive qrel. This gives us highly incomplete qrels, as noted in the original description \citep{bajaj2016ms}. We should in no way expect the positive qrel to be the ``best answer''. We found that training and evaluating using these sparse qrels gives us results that are quite correlated with results using much more comprehensive NIST judgments \citep{craswell2019overview, craswell2020overview}. Further study is needed to understand why this works, but we suspect it's important that the qrel is selected from a Bing ranking that has access to information that's unavailable to TREC participants, such as billions of past queries. This means the selected qrel is not biased towards some existing academic approach such as BM25. For each query that has a qrel, we generated a BM25 top-1000 for use in a reranking task and also allowed fullrank from the 8.8 million passages. We used the same split as in the NLG task: training ($80\%$), dev ($10\%$) and eval ($10\%$).

To generate the v1 document ranking data, we collected the corresponding urls for which the passages were extracted. Using these 3.5 million URLS, we obtained the associated document title and body corresponding to the ranking qrels. It is worth noting that the original passages were extracted between January 2016 and February 18 while the full documents were extracted in March of 2018 and as a result only 3.2 million URLs were still in existence. From these documents, the body text had the HTML removed and focused on the main content of the page, removing web-page boilerplate such as navigation menus. Since we extracted the document text more than a year later than the passage data and used a completely different document parsing and processing pipeline (which unfortunately had character set processing issues) there was a chance that some pages that had a relevant passage no longer existed, no longer contained the passage, or even had the section of text with the passage accidentally removed as boilerplate. These are all realistic things to happen in a real-world application, where the document corpus is constantly changing, we do not wish to throw away our old relevance labels, and indeed we may not have budget to generate new labels. Doing a better job of generating a clean dataset using old labels is what we have now done in generating the v2 data. Qrels for the document task were assigned by assuming that a relevant passage qrel transfers to the document level as a positive document qrel. We generated top-100 document rankings using Indri, for use in a reranking task and also allowed fullrank from the 3.2 million documents.

The v1 data had several problems. The corpus was generated based on the queries, such that each passage and each document is in the corpus due to one of our million original queries. For each document in the corpus there may only be one passage in the passage dataset (and on average 2.8 passages per document), but that passage was identified by Bing in relation to one of the MS MARCO queries, possibly a test query. This is unrealistic, since a real system would be able to generate many candidate passages per document, and would not know what the test queries will be ahead of time. Therefore, we had to forbid participants from considering the passage-document mapping. The document dataset had several problems with character sets and missing whitespaces.

\paragraph{MS MARCO ranking v2 datasets.}
The MS MARCO passage and document ranking v2 datasets have been used for the first time in TREC 2021.

The v2 data starts by identifying documents. Of the original 3.2 million MS MARCO documents, we were able to still find content for 2.7 million on the Web. We added an additional 9.2 million documents, selected to be the kind of documents that had useful passages of text in past Bing queries, giving a total of 11.9 million documents. For each document we ran a query-independent proprietary algorithm for identifying promising passages, and selected the best non-overlapping passages, giving on average 11.6 passages per document. This gives us our 138 million passages in the v2 passage corpus. We mapped the document qrels at the URL level, for training, dev and eval. The chance that the document is no longer relevant to the query, which also was a concern in v1 data, is now increased since the document content was extracted at a later date. We can consider how big this problem is by analyzing the disagreement rate between MS MARCO qrels and NIST qrels (in v1 and v2), and seeing whether training on MS MARCO qrels yields improved NIST NDCG on the test set. For mapping passage qrels, we required that the passage comes from the same URL as the original passage, and has sufficient text similarity to the positive passage text from v1. 

It is now possible for participants to use the passage-document mapping in participation, for example by considering document information in passage ranking, passage information in document ranking, and so on. Using a larger corpus prevents participants from proposing completely unscalable ranking approaches. The new dataset has fewer character encoding and whitespace issues, and could form the basis for future tasks that include some elements of additional document processing, such as extracting even shorter (phrase) answers.

\section{Results and analysis}
\label{sec:result}

\paragraph{Submitted runs}
A total of $19$ groups participated in the TREC 2021 Deep Learning Track.
Among them, $11$ groups participated in both the document and the passage ranking tasks, and of the remaining four groups participated only in the document ranking task and another four groups on in passage ranking task.
We also solicited baseline runs to enrich the judgment pools which are reported under a separate ``BASELINES'' group as we did in previous years of the track.
Across all groups, we received a total of $129$ run submissions, including $66$ document ranking runs and $63$ passage ranking runs.
This also includes $37$ baseline runs---$20$ for document ranking and $17$ for passage ranking.
Unlike previous years, this year some of the baseline runs also employed neural methods.
Table~\ref{tbl:runs-by-type} summarizes the submissions statistics for this year's track.

\begin{table}
    \centering
    \caption{TREC 2021 Deep Learning Track run submission statistics.}
    \begin{tabular}{lrr}
    \hline
    \hline
        & \textbf{Document ranking} & \textbf{Passage ranking} \\
        \hline
        Number of groups & 15 & 15 \\
        Number of total runs & 66 & 63 \\
        Number of baseline runs & 20 & 17 \\
        Number of runs w/ category: nnlm & 48 & 50 \\
        Number of runs w/ category: nn & 0 & 0 \\
        Number of runs w/ category: trad & 18 & 13 \\
        Number of runs w/ category: rerank & 15 & 12 \\
        Number of runs w/ category: fullrank & 51 & 51 \\
        \hline
        \hline
    \end{tabular}
    \label{tbl:runs-by-type}
\end{table}

This year we had fewer participating groups (19 groups) compared to last year (25 groups) but more than the inaugural year of the track (15 groups) likely due to the delay in releasing the v2 collection.
However, we received a larger number of runs this year compared to previous years (75 runs in 2019 and 123 runs in 2020), although a larger number of baseline runs contributed towards that growth.

This year we asked participants to self-classify each of their runs under the following three categories (same taxonomy as was employed in our previous track overview papers~\citep{craswell2019overview, craswell2020overview}):
\begin{itemize}
    \item trad: No neural representation learning---\eg, classical learning to rank, PRF, and BM25
    \item nn: Representation learning with text as input, but not using a pre-trained model
    \item nnlm: Using a pre-trained model in any part of the pipeline---\eg, neural document expansion and BERT-style reranking
\end{itemize}

The largest category of runs was of type ``nnlm'' constituting $76\%$ of submissions across both tasks this year.
This was a significant increase over previous years---$44\%$ in 2019 and $57\%$ in 2020---while the percentage of ``trad'' runs have remained relatively stable over the years---$29\%$ in 2019, $33\%$ in 2020, and $24\%$ in 2021.
A significant shift also happened for the ``nn'' category over the years, decreasing from $27\%$ in 2019 to $10\%$ in 2020 and altogether disappearing as a category this year.
This may reflect a convergence in the neural IR community, and the IR community in general, towards large language models, although whether this homogenization of approaches is healthy or premature is yet to be seen.

Participants were also asked to categorize their runs based on subtasks:
\begin{itemize}
    \item Rerank: Reranking the official top-100 candidates
    \item Fullrank: Full ranking from the collection (retrieval)
\end{itemize}

We observed an increase in the percentage of ``fullrank'' runs this year---$79\%$ compared to $72\%$ in 2019 and $70\%$ in 2020.
The biggest increase in ``fullrank'' runs this year were for the passage ranking task---$81\%$ this year compared to $70\%$ in 2019 and $69\%$ in 2020---which may have been partially influenced by the reduction in size of the official reranking candidate set for the passage ranking task from $1000$ (as in previous years) to $100$ this year.
The growing percentage of ``fullrank'' runs may also be due to increasing application of neural methods in the full retrieval setting---either using dense retrieval methods~\citep{lee2019latent} or query term independent neural ranking models~\citep{mitra2019incorporating}.
Coincidentally, this year, we also asked participants to tell us
\begin{enumerate*}[label=(\roman*)]
    \item if their runs employed dense retrieval methods, and
    \item if the retrieval was performed in a single-stage under full retrieval setting.
\end{enumerate*}

\paragraph{Overall results}
Table~\ref{tab:document_ranking} and Table~\ref{tab:passage_ranking} presents a standard set of relevance quality metrics for document and passage ranking runs, respectively, as we have reported for the track in previous years.
We removed couple of runs with very low metric values, which may have been due to some issue with run generation, from our result tables and analysis plots.
The reported metrics include Reciprocal Rank (RR)~\citep{craswell2009mean}, Normalized Discounted Cumulative Gains (NDCG)~\citep{JK2002}, Normalized Cumulative Gains (NCG)~\citep{rosset2018optimizing}, and Average Precision (AP)~\citep{zhu2004recall} computed using the NIST judgments.
In addition, we also report RR computed based on the original sparse MS MARCO labels.

In subsequent discussions, we employ NDCG@10 as our primary evaluation metric to analyze ranking quality produced by different methods.
To analyze how different approaches compare beyond just the relevance of top-ranked results, we use NCG@100 which correlates more with how often relevant results are in the top-100 candidate set even if they are not eventually ranked as highly as appropriate.
We employ RR mostly for comparison between NIST and MS MARCO labels as the latter consists of binary judgments.

\begin{table}[]
\caption{Summary of results for document ranking runs.}
\scriptsize
\centering
\begin{tabular}{lllllllrrlr}
\toprule
{} &             group &   subtask & neural &   stage & dense ret. & RR (MS) &      RR &  NDCG@10 & NCG@100 &      AP \\
\midrule
pash\_doc\_f1     &              PASH &  fullrank &   nnlm &   multi &        yes &  0.3027 &  0.9795 &   0.7437 &  0.5037 &  0.3111 \\
pash\_doc\_f4     &              PASH &  fullrank &   nnlm &   multi &        yes &  0.2999 &  0.9795 &   0.7404 &  0.5983 &  0.3498 \\
pash\_doc\_f5     &              PASH &  fullrank &   nnlm &   multi &        yes &  0.3018 &  0.9795 &   0.7368 &  0.5992 &  0.3521 \\
d\_f10\_mdt53b    &            h2oloo &  fullrank &   nnlm &   multi &        yes &  0.4297 &  0.9883 &   0.7256 &  0.5162 &  0.2837 \\
NLE\_D\_v1        &               NLE &  fullrank &   nnlm &   multi &         no &  0.2536 &  1.0000 &   0.7215 &  0.5564 &  0.3133 \\
uogTrDDQt5      &             uogTr &  fullrank &   nnlm &   multi &         no &  0.3044 &  0.9737 &   0.7201 &  0.4849 &  0.2963 \\
pash\_doc\_r3     &              PASH &    rerank &   nnlm &   multi &         no &  0.3022 &  0.9772 &   0.7164 &  0.4376 &  0.2672 \\
pash\_doc\_r1     &              PASH &    rerank &   nnlm &   multi &         no &  0.3362 &  0.9772 &   0.7150 &  0.4376 &  0.2665 \\
pash\_doc\_r2     &              PASH &    rerank &   nnlm &   multi &         no &  0.3198 &  0.9772 &   0.7076 &  0.4376 &  0.2640 \\
d\_f10\_mt53b     &            h2oloo &  fullrank &   nnlm &   multi &        yes &  0.4009 &  0.9605 &   0.7024 &  0.5162 &  0.2767 \\
d\_fusion00      &         BASELINES &  fullrank &   nnlm &  single &        yes &  0.2451 &  0.9630 &   0.7003 &  0.5226 &  0.2887 \\
uogTrBaseDDQC   &         BASELINES &  fullrank &   nnlm &   multi &         no &  0.2663 &  0.9649 &   0.6966 &  0.4849 &  0.2869 \\
NLE\_D\_V1andV2   &               NLE &  fullrank &   nnlm &   multi &         no &  0.2697 &  0.9503 &   0.6871 &  0.5372 &  0.2969 \\
d\_fusion10      &         BASELINES &  fullrank &   nnlm &  single &        yes &  0.2951 &  0.9423 &   0.6831 &  0.4752 &  0.2518 \\
bcai\_bertm1\_ens &              bcai &  fullrank &   nnlm &   multi &        yes &  0.4793 &  0.9444 &   0.6812 &  0.4974 &  0.2559 \\
CIP\_run2        &               CIP &    rerank &   nnlm &   multi &         no &  0.3429 &  0.9373 &   0.6783 &  0.4376 &  0.2478 \\
bl\_bcai\_wloo\_d  &         BASELINES &  fullrank &   nnlm &   multi &        yes &  0.4681 &  0.9230 &   0.6762 &  0.4752 &  0.2534 \\
CIP\_run1        &               CIP &    rerank &   nnlm &   multi &         no &  0.3668 &  0.9505 &   0.6755 &  0.4376 &  0.2445 \\
max-firstp-pass &         CFDA\_CLIP &  fullrank &   nnlm &  single &        yes &  0.3015 &  0.9363 &   0.6727 &  0.4383 &  0.2321 \\
maxp-firstp     &         CFDA\_CLIP &  fullrank &   nnlm &  single &        yes &  0.2642 &  0.9436 &   0.6709 &  0.4308 &  0.2278 \\
parade\_bm25     &              mpii &  fullrank &   nnlm &   multi &         no &  0.3395 &  0.9592 &   0.6707 &  0.4376 &  0.2490 \\
maxp            &         CFDA\_CLIP &  fullrank &   nnlm &  single &        yes &  0.2535 &  0.9538 &   0.6672 &  0.4306 &  0.2308 \\
bl\_bcai\_nn\_rtr  &         BASELINES &  fullrank &   nnlm &  single &        yes &  0.2865 &  0.9630 &   0.6671 &  0.5044 &  0.2749 \\
CIP\_run3        &               CIP &    rerank &   nnlm &   multi &         no &  0.3350 &  0.9567 &   0.6668 &  0.4376 &  0.2457 \\
d\_f10\_mdt5base  &            h2oloo &  fullrank &   nnlm &   multi &        yes &  0.4198 &  0.9401 &   0.6606 &  0.4758 &  0.2376 \\
d\_tct0          &         BASELINES &  fullrank &   nnlm &  single &        yes &  0.2203 &  0.9455 &   0.6537 &  0.4562 &  0.2418 \\
ielab-roberta1d &         BASELINES &  fullrank &   nnlm &   multi &         no &  0.2137 &  0.9591 &   0.6522 &  0.3484 &  0.2039 \\
TUW\_IDCM\_S4     &         TU\_Vienna &    rerank &   nnlm &   multi &         no &  0.3325 &  0.9115 &   0.6494 &  0.4376 &  0.2460 \\
watdrf          &  Waterloo\_Cormack &    rerank &   trad &   multi &        yes &  0.1736 &  0.9693 &   0.6467 &  0.4218 &  0.2533 \\
TUW\_IDCM\_ALL    &         TU\_Vienna &    rerank &   nnlm &   multi &         no &  0.3461 &  0.9092 &   0.6455 &  0.4376 &  0.2460 \\
ielab-roberta2d &         BASELINES &  fullrank &   nnlm &   multi &         no &  0.2624 &  0.9470 &   0.6431 &  0.3484 &  0.2040 \\
ielab-AD-uni-d  &             ielab &  fullrank &   nnlm &  single &        yes &  0.2377 &  0.9684 &   0.6424 &  0.4754 &  0.2492 \\
ielab-uniCOIL-d &             ielab &  fullrank &   nnlm &  single &         no &  0.2377 &  0.9684 &   0.6424 &  0.4754 &  0.2492 \\
doc\_full\_100    &           ALIBABA &  fullrank &   nnlm &   multi &        yes &  0.4022 &  0.9308 &   0.6414 &  0.5021 &  0.2465 \\
doc\_full\_100e   &           ALIBABA &  fullrank &   nnlm &   multi &        yes &  0.4022 &  0.9308 &   0.6414 &  0.5021 &  0.2465 \\
uogTrDot5pmp    &         BASELINES &  fullrank &   nnlm &   multi &        yes &  0.2152 &  0.9386 &   0.6411 &  0.3540 &  0.2077 \\
watdrd          &  Waterloo\_Cormack &    rerank &   trad &   multi &        yes &  0.1737 &  0.9520 &   0.6411 &  0.4376 &  0.2529 \\
Fast\_Forward\_2  &               L3S &  fullrank &   nnlm &   multi &         no &  0.1959 &  0.9247 &   0.6338 &  0.5279 &  0.2773 \\
d\_unicoil0      &         BASELINES &  fullrank &   nnlm &  single &         no &  0.1967 &  0.9088 &   0.6325 &  0.4784 &  0.2495 \\
watdrp          &  Waterloo\_Cormack &    rerank &   trad &   multi &        yes &  0.1818 &  0.9386 &   0.6307 &  0.4218 &  0.2419 \\
parade\_h3       &              mpii &  fullrank &   nnlm &   multi &         no &  0.3064 &  0.9488 &   0.6295 &  0.5058 &  0.2590 \\
Fast\_Forward\_5  &               L3S &  fullrank &   nnlm &   multi &         no &  0.2165 &  0.9373 &   0.6282 &  0.5143 &  0.2730 \\
d\_tct1          &         BASELINES &  fullrank &   nnlm &  single &        yes &  0.2754 &  0.9190 &   0.6269 &  0.3568 &  0.1794 \\
dseg\_bm25rm3    &         BASELINES &  fullrank &   trad &  single &         no &  0.1582 &  0.9018 &   0.6185 &  0.5232 &  0.2933 \\
doc\_rank\_100    &           ALIBABA &    rerank &   nnlm &   multi &        yes &  0.3835 &  0.9392 &   0.6175 &  0.4376 &  0.2168 \\
bl\_bcai\_trad    &         BASELINES &  fullrank &   trad &   multi &         no &  0.2083 &  0.9276 &   0.6136 &  0.4735 &  0.2494 \\
watdff          &  Waterloo\_Cormack &  fullrank &   trad &  single &        yes &  0.1164 &  0.9211 &   0.6060 &  0.5319 &  0.2961 \\
ielab-TILDEv2d  &             ielab &  fullrank &   nnlm &   multi &         no &  0.1935 &  0.8829 &   0.6022 &  0.4335 &  0.2262 \\
maxp\_h3         &              mpii &  fullrank &   nnlm &   multi &         no &  0.2772 &  0.9313 &   0.6017 &  0.4596 &  0.2198 \\
NLE\_D\_quick     &               NLE &  fullrank &   nnlm &  single &         no &  0.1816 &  0.9024 &   0.6015 &  0.4344 &  0.2198 \\
bigrams\_cont\_qe &     CERTH\_ITI\_M4D &  fullrank &   nnlm &  single &         no &  0.1488 &  0.8751 &   0.5941 &  0.5178 &  0.2957 \\
bigram\_qe\_cedr  &     CERTH\_ITI\_M4D &  fullrank &   nnlm &   multi &        yes &  0.1730 &  0.8872 &   0.5920 &  0.5273 &  0.2802 \\
webis-dl-3      &             Webis &    rerank &   trad &   multi &         no &  0.3205 &  0.9488 &   0.5918 &  0.4376 &  0.2305 \\
Fast\_Forward\_7  &               L3S &  fullrank &   nnlm &   multi &         no &  0.2180 &  0.8952 &   0.5905 &  0.4880 &  0.2525 \\
webis-dl-1      &             Webis &    rerank &   trad &   multi &         no &  0.3409 &  0.9356 &   0.5831 &  0.4376 &  0.2254 \\
dseg\_bm25       &         BASELINES &  fullrank &   trad &  single &         no &  0.2066 &  0.8937 &   0.5776 &  0.4709 &  0.2436 \\
webis-dl-2      &             Webis &    rerank &   trad &   multi &         no &  0.2693 &  0.9396 &   0.5747 &  0.4376 &  0.2224 \\
uogTrBaseDD     &         BASELINES &  fullrank &   trad &  single &         no &  0.2019 &  0.8297 &   0.5704 &  0.4849 &  0.2487 \\
uogTrBaseDDQ    &         BASELINES &  fullrank &   trad &  single &         no &  0.2019 &  0.8297 &   0.5704 &  0.4849 &  0.2487 \\
watdfd          &  Waterloo\_Cormack &  fullrank &   trad &  single &        yes &  0.1189 &  0.8895 &   0.5615 &  0.4833 &  0.2504 \\
watdfp          &  Waterloo\_Cormack &  fullrank &   trad &  single &        yes &  0.1263 &  0.8874 &   0.5580 &  0.4561 &  0.2351 \\
d\_bm25rm3       &         BASELINES &  fullrank &   trad &  single &         no &  0.1240 &  0.7994 &   0.5339 &  0.4614 &  0.2453 \\
d\_bm25          &         BASELINES &  fullrank &   trad &  single &         no &  0.1647 &  0.8367 &   0.5116 &  0.4376 &  0.2126 \\
uogTrBaseDDQpmp &         BASELINES &  fullrank &   trad &  single &         no &  0.1795 &  0.8316 &   0.5105 &  0.3972 &  0.2029 \\
uogTrBaseDDpmp  &         BASELINES &  fullrank &   trad &  single &         no &  0.1788 &  0.8563 &   0.5070 &  0.3594 &  0.1769 \\
\bottomrule
\end{tabular}
\label{tab:document_ranking}
\end{table}

\begin{table}[]
\caption{Summary of results for passage ranking runs.}
\scriptsize
\centering
\begin{tabular}{lllllllrrlr}
\toprule
{} &             group &   subtask & neural &   stage & dense ret. & RR (MS) &      RR &  NDCG@10 & NCG@100 &      AP \\
\midrule
pash\_f1         &              PASH &  fullrank &   nnlm &   multi &        yes &  0.2058 &  0.8732 &   0.7494 &  0.5249 &  0.3193 \\
pash\_f2         &              PASH &  fullrank &   nnlm &   multi &        yes &  0.2068 &  0.8737 &   0.7494 &  0.5721 &  0.3318 \\
pash\_f3         &              PASH &  fullrank &   nnlm &   multi &        yes &  0.2068 &  0.8737 &   0.7494 &  0.5882 &  0.3378 \\
NLE\_P\_v1        &               NLE &  fullrank &   nnlm &   multi &         no &  0.2079 &  0.8697 &   0.7347 &  0.6172 &  0.3923 \\
pash\_r2         &              PASH &    rerank &   nnlm &   multi &         no &  0.1939 &  0.8678 &   0.7076 &  0.3862 &  0.2389 \\
pash\_r3         &              PASH &    rerank &   nnlm &   multi &        yes &  0.1939 &  0.8675 &   0.7072 &  0.3862 &  0.2385 \\
yorku21\_a       &             yorku &  fullrank &   nnlm &   multi &        yes &  0.1903 &  0.8629 &   0.6965 &  0.5456 &  0.3309 \\
pash\_r1         &              PASH &    rerank &   nnlm &   multi &         no &  0.2359 &  0.8663 &   0.6951 &  0.3862 &  0.2362 \\
yorku21\_c       &             yorku &  fullrank &   nnlm &   multi &        yes &  0.1794 &  0.8393 &   0.6930 &  0.5413 &  0.3323 \\
mono\_electra\_h3 &              mpii &  fullrank &   nnlm &   multi &         no &  0.2538 &  0.7887 &   0.6753 &  0.5286 &  0.2880 \\
NLE\_P\_V1andV2   &               NLE &  fullrank &   nnlm &   multi &         no &  0.2127 &  0.7756 &   0.6724 &  0.5834 &  0.3381 \\
ielab-AD-uni    &             ielab &  fullrank &   nnlm &  single &        yes &  0.1618 &  0.8045 &   0.6714 &  0.5239 &  0.2842 \\
p\_f10\_mdt53b    &            h2oloo &  fullrank &   nnlm &   multi &        yes &  0.3728 &  0.7871 &   0.6617 &  0.3961 &  0.2089 \\
ielab-robertav1 &         BASELINES &  fullrank &   nnlm &   multi &         no &  0.1851 &  0.8144 &   0.6551 &  0.3862 &  0.2246 \\
uogTrPot5       &         BASELINES &  fullrank &   nnlm &   multi &        yes &  0.1931 &  0.8160 &   0.6517 &  0.3980 &  0.2499 \\
bcai\_p\_vbert    &              bcai &  fullrank &   nnlm &   multi &        yes &  0.2846 &  0.7957 &   0.6502 &  0.4296 &  0.2323 \\
ihsm\_colbert64  &              IHSM &    rerank &   nnlm &   multi &         no &  0.1498 &  0.8463 &   0.6453 &  0.3862 &  0.2148 \\
pass\_full\_1000  &           ALIBABA &  fullrank &   nnlm &   multi &        yes &  0.2830 &  0.7695 &   0.6434 &  0.4238 &  0.2413 \\
ielab-uniCOIL   &             ielab &  fullrank &   nnlm &  single &         no &  0.1873 &  0.7975 &   0.6420 &  0.5207 &  0.2745 \\
p\_f10\_mt53b     &            h2oloo &  fullrank &   nnlm &   multi &        yes &  0.3125 &  0.7700 &   0.6394 &  0.3961 &  0.1992 \\
ihsm\_bicolbert  &              IHSM &    rerank &   nnlm &   multi &         no &  0.1868 &  0.7962 &   0.6393 &  0.3862 &  0.2111 \\
bcai\_p\_mbert    &              bcai &  fullrank &   nnlm &   multi &        yes &  0.2902 &  0.7765 &   0.6363 &  0.4600 &  0.2487 \\
ihsm\_poly8q     &              IHSM &    rerank &   nnlm &   multi &         no &  0.1468 &  0.8233 &   0.6342 &  0.3862 &  0.2059 \\
ielab-robertav2 &         BASELINES &  fullrank &   nnlm &   multi &         no &  0.2231 &  0.7645 &   0.6226 &  0.3862 &  0.2000 \\
p\_f10\_mdt5base  &            h2oloo &  fullrank &   nnlm &   multi &        yes &  0.3360 &  0.7486 &   0.6193 &  0.3711 &  0.1819 \\
mono\_h3         &              mpii &  fullrank &   nnlm &   multi &         no &  0.2455 &  0.7267 &   0.6115 &  0.4964 &  0.2548 \\
NLE\_P\_quick     &               NLE &  fullrank &   nnlm &  single &         no &  0.1141 &  0.7342 &   0.6087 &  0.4884 &  0.2443 \\
pass\_full\_1000e &           ALIBABA &  fullrank &   nnlm &   multi &        yes &  0.2772 &  0.7339 &   0.6071 &  0.4676 &  0.2425 \\
mono\_d3         &              mpii &  fullrank &   nnlm &   multi &         no &  0.2493 &  0.7341 &   0.6037 &  0.4051 &  0.2150 \\
bl\_bcai\_wloo\_p  &         BASELINES &  fullrank &   nnlm &   multi &        yes &  0.2108 &  0.7691 &   0.6029 &  0.4098 &  0.2077 \\
p\_fusion10      &         BASELINES &  fullrank &   nnlm &  single &        yes &  0.1776 &  0.7854 &   0.5857 &  0.4098 &  0.1855 \\
ielab-TILDEv2   &             ielab &  fullrank &   nnlm &   multi &         no &  0.1489 &  0.6926 &   0.5825 &  0.4511 &  0.2112 \\
p\_unicoil0      &         BASELINES &  fullrank &   nnlm &  single &         no &  0.1333 &  0.6788 &   0.5785 &  0.4408 &  0.2165 \\
p\_fusion00      &         BASELINES &  fullrank &   nnlm &  single &        yes &  0.1706 &  0.7783 &   0.5713 &  0.4675 &  0.2005 \\
yorku21\_b       &             yorku &  fullrank &   nnlm &  single &        yes &  0.1099 &  0.7180 &   0.5694 &  0.4112 &  0.2032 \\
TUW\_TAS-B\_768   &         TU\_Vienna &  fullrank &   nnlm &  single &        yes &  0.0723 &  0.7333 &   0.5619 &  0.4666 &  0.2093 \\
Fast\_ForwardP\_2 &               L3S &  fullrank &   nnlm &   multi &         no &  0.1754 &  0.6486 &   0.5521 &  0.4579 &  0.1998 \\
watprp          &  Waterloo\_Cormack &    rerank &   trad &   multi &        yes &  0.1127 &  0.6827 &   0.5493 &  0.3862 &  0.1728 \\
Fast\_Forward\_3  &               L3S &  fullrank &   nnlm &   multi &         no &  0.1644 &  0.6401 &   0.5451 &  0.4547 &  0.1927 \\
TUW\_TAS-B\_ANN   &         TU\_Vienna &  fullrank &   nnlm &  single &        yes &  0.0842 &  0.7015 &   0.5426 &  0.4595 &  0.1932 \\
pass\_rank\_100   &           ALIBABA &    rerank &   nnlm &   multi &        yes &  0.2161 &  0.6588 &   0.5389 &  0.3862 &  0.1781 \\
bl\_bcai\_p\_nn\_rt &         BASELINES &  fullrank &   nnlm &   multi &        yes &  0.1844 &  0.6852 &   0.5245 &  0.3926 &  0.1691 \\
watprf          &  Waterloo\_Cormack &    rerank &   trad &   multi &        yes &  0.1051 &  0.6050 &   0.5184 &  0.3862 &  0.1631 \\
Fast\_ForwardP\_5 &               L3S &  fullrank &   nnlm &   multi &         no &  0.1482 &  0.6034 &   0.5132 &  0.4294 &  0.1728 \\
top1000         &        UAmsterdam &  fullrank &   nnlm &   multi &        yes &  0.0411 &  0.6566 &   0.5104 &  0.2387 &  0.1217 \\
p\_tct0          &         BASELINES &  fullrank &   nnlm &  single &        yes &  0.1780 &  0.6574 &   0.5001 &  0.3552 &  0.1332 \\
TUW\_DR\_Base     &         TU\_Vienna &  fullrank &   nnlm &  single &        yes &  0.0956 &  0.6768 &   0.4991 &  0.3675 &  0.1540 \\
watpfp          &  Waterloo\_Cormack &  fullrank &   trad &  single &        yes &  0.0934 &  0.5985 &   0.4950 &  0.4044 &  0.1738 \\
uogTrBasePDQ    &         BASELINES &  fullrank &   trad &  single &         no &  0.1602 &  0.5611 &   0.4747 &  0.3980 &  0.1724 \\
watprd          &  Waterloo\_Cormack &    rerank &   trad &   multi &        yes &  0.1163 &  0.6240 &   0.4698 &  0.3862 &  0.1442 \\
uogTrBasePD     &         BASELINES &  fullrank &   trad &  single &         no &  0.1617 &  0.5610 &   0.4619 &  0.3642 &  0.1439 \\
uogTrPC         &             uogTr &  fullrank &   nnlm &   multi &        yes &  0.0733 &  0.6119 &   0.4611 &  0.1295 &  0.0855 \\
p\_tct1          &         BASELINES &  fullrank &   nnlm &  single &        yes &  0.1550 &  0.5703 &   0.4499 &  0.3118 &  0.1159 \\
p\_bm25rm3       &         BASELINES &  fullrank &   trad &  single &         no &  0.1371 &  0.4925 &   0.4480 &  0.3989 &  0.1632 \\
p\_bm25          &         BASELINES &  fullrank &   trad &  single &         no &  0.1277 &  0.5060 &   0.4458 &  0.3862 &  0.1357 \\
WLUPassage      &               WLU &    rerank &   nnlm &   multi &         no &  0.0909 &  0.5682 &   0.4432 &  0.3862 &  0.1348 \\
watpff          &  Waterloo\_Cormack &  fullrank &   trad &  single &        yes &  0.0722 &  0.5104 &   0.4408 &  0.3915 &  0.1346 \\
bl\_bcai\_p\_trad  &         BASELINES &  fullrank &   trad &   multi &         no &  0.1493 &  0.5086 &   0.4261 &  0.3435 &  0.1133 \\
WLUPassage1     &               WLU &    rerank &   nnlm &   multi &         no &  0.0453 &  0.4886 &   0.4093 &  0.3862 &  0.1170 \\
paug\_bm25       &         BASELINES &  fullrank &   trad &  single &         no &  0.0848 &  0.5303 &   0.3977 &  0.3116 &  0.0977 \\
paug\_bm25rm3    &         BASELINES &  fullrank &   trad &  single &         no &  0.0676 &  0.4906 &   0.3906 &  0.3152 &  0.1050 \\
watpfd          &  Waterloo\_Cormack &  fullrank &   trad &  single &        yes &  0.0690 &  0.4833 &   0.3672 &  0.2535 &  0.0688 \\
\bottomrule
\end{tabular}
\label{tab:passage_ranking}
\end{table}

\begin{figure}
  \center
  \begin{subfigure}{.49\textwidth}
    \includegraphics[width=\textwidth]{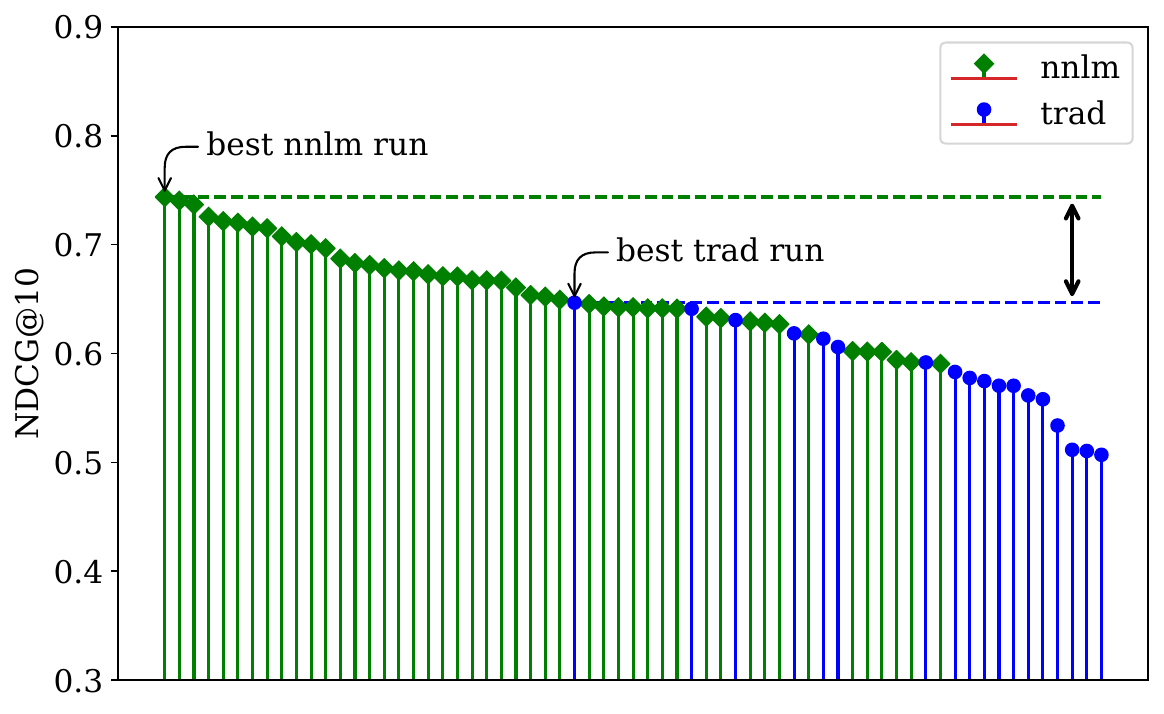}
    \caption{Document retrieval task}
    \label{fig:model-task-docs-stem-by-model-type}
  \end{subfigure}
  \hfill
  \begin{subfigure}{.49\textwidth}
    \includegraphics[width=\textwidth]{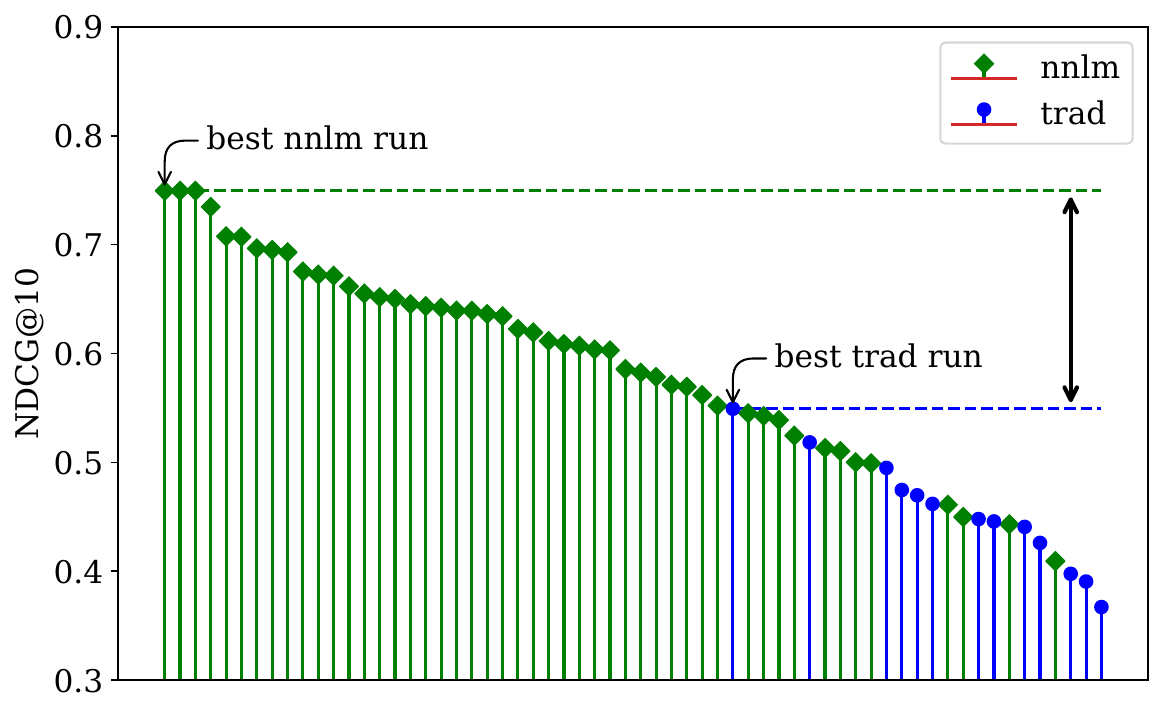}
    \caption{Passage retrieval task}
    \label{fig:model-task-passages-stem-by-model-type}
  \end{subfigure}
  \caption{NDCG@10 results by run type. As in the previous two years, ``nnlm'' runs continue to outperform over ``trad'' runs for both tasks.}
  \label{fig:model-stem-by-model-type}
\end{figure}

\paragraph{Neural \vs traditional methods.}
Figure~\ref{fig:model-stem-by-model-type} summarizes the evaluation results by run type---\ie, comparing ``nnlm'' \vs ``trad'' runs.
Across both document and passage ranking tasks, ``nnlm'' runs continue to significantly outperform ``trad'' runs this year.
For the document ranking task, the best performing ``nnlm'' run improves NDCG@10 over the best performing ``trad'' run by $15\%$ this year, compared to $29\%$ in 2019 and $23\%$ in 2020.
On the other hand, for the passage ranking task, the NDCG@10 gap between the best performing run in `nnlm'' and ``trad'' categories is $36\%$, while the same was $38\%$ in 2019 and $42\%$ in 2020.
Comparing percentage improvements across different year's tracks or across different tasks in the same year is not very meaningful due to differences in underlying data distributions.
However, we still find it interesting that the percentage improvements this year are lower than previous years, and that the gap has been consistently bigger for the passage ranking task compared to the document ranking task for each year of the track.

\begin{figure}
\includegraphics[width=\textwidth]{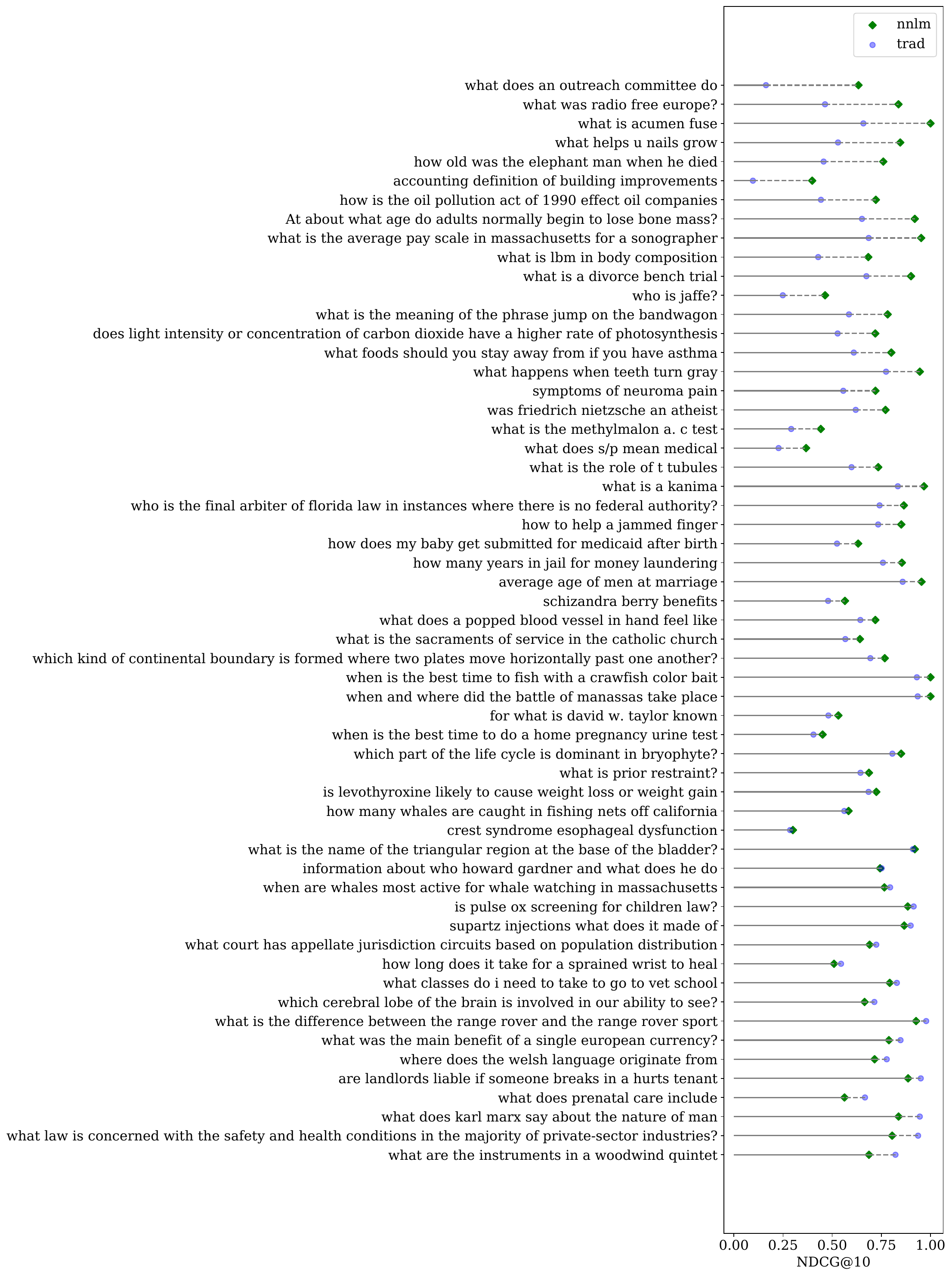}
\caption{Comparison of the best ``nnlm'' and ``trad'' runs on individual test queries for the document retrieval task. Queries are sorted by difference in mean performance between ``nnlm'' and ``trad'' runs. Queries on which ``nnlm'' wins with large margin are at the top.}
\label{fig:model-task-docs-bar-per-query}
\end{figure}

\begin{figure}
\includegraphics[width=\textwidth]{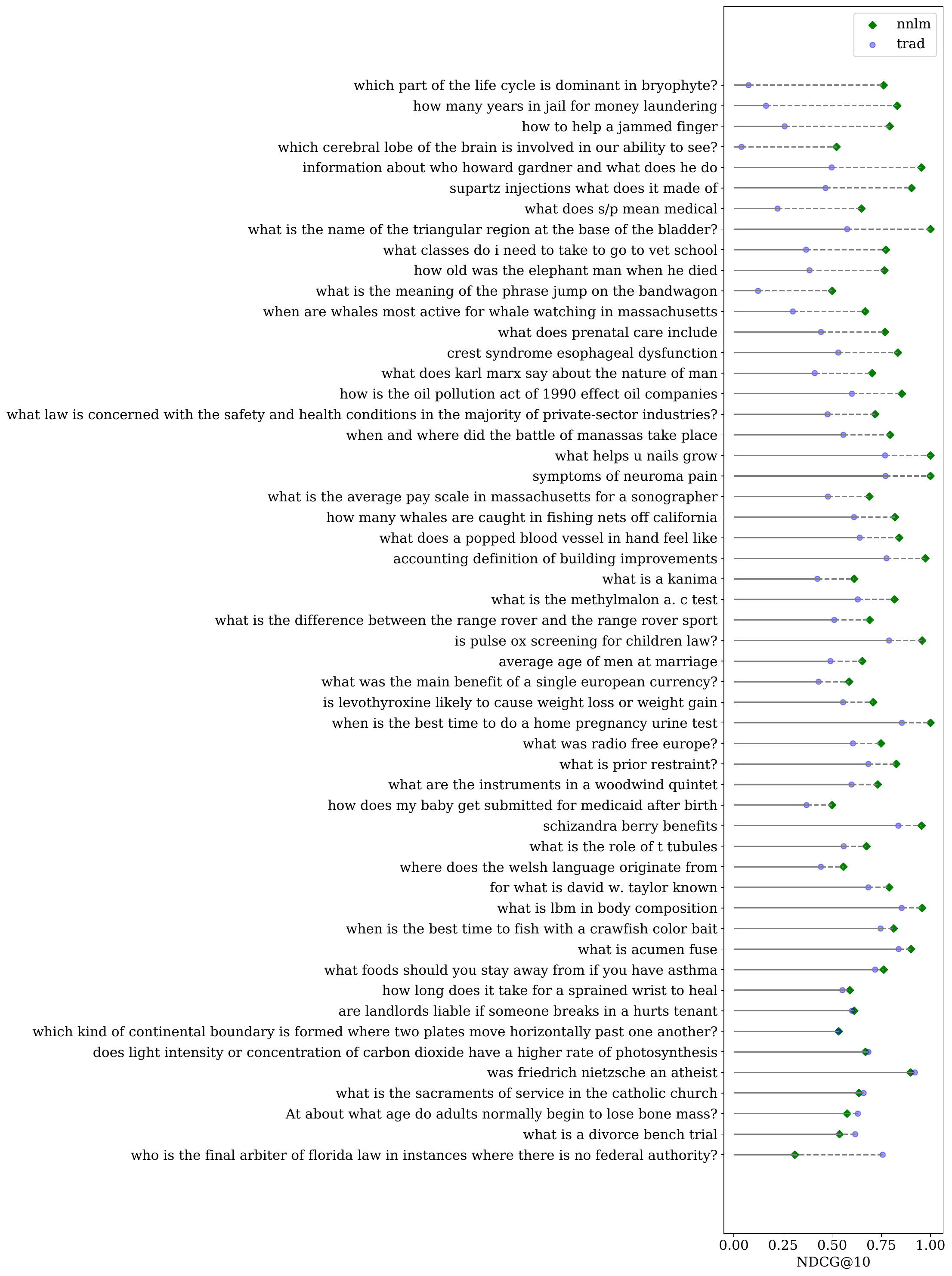}
\caption{Comparison of the best ``nnlm'' and ``trad'' runs on individual test queries for the passage retrieval task. Queries are sorted by difference in mean performance between ``nnlm'' and ``trad'' runs. Queries on which ``nnlm'' wins with large margin are at the top.}
\label{fig:model-task-passages-bar-per-query}
\end{figure}

Figure~\ref{fig:model-task-docs-bar-per-query} and \ref{fig:model-task-passages-bar-per-query} shows a query-level comparison between the best ``nnlm'' and ``trad'' runs for the document and the passage ranking tasks, respectively.
The best ``nnlm'' run outperforms the best ``trad run'' on 41 out of 57 ($72\%$) queries for the document ranking task---a drop-off from $84\%$ as in the previous two years.
For the passage ranking task, the best ``nnlm'' run wins on 47 out of 53 ($89\%$) queries against the best ``trad'' run, which is marginally higher than $84\%$ in 2019 and $88\%$ in 2020.

\begin{figure}
  \center
  \begin{subfigure}{.49\textwidth}
    \includegraphics[width=\textwidth]{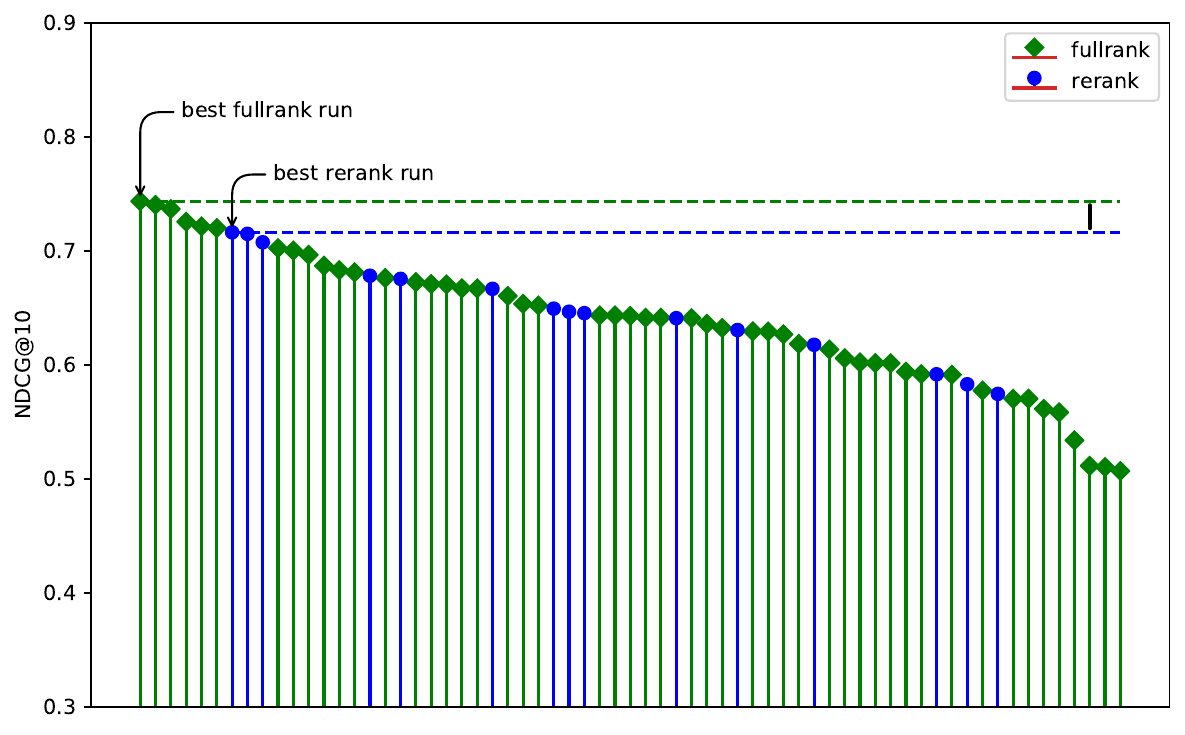}
    \caption{NDCG@10 for runs on the document retrieval task}
    \label{fig:model-task-docs-stem-by-subtask}
  \end{subfigure}
  \hfill
  \begin{subfigure}{.49\textwidth}
    \includegraphics[width=\textwidth]{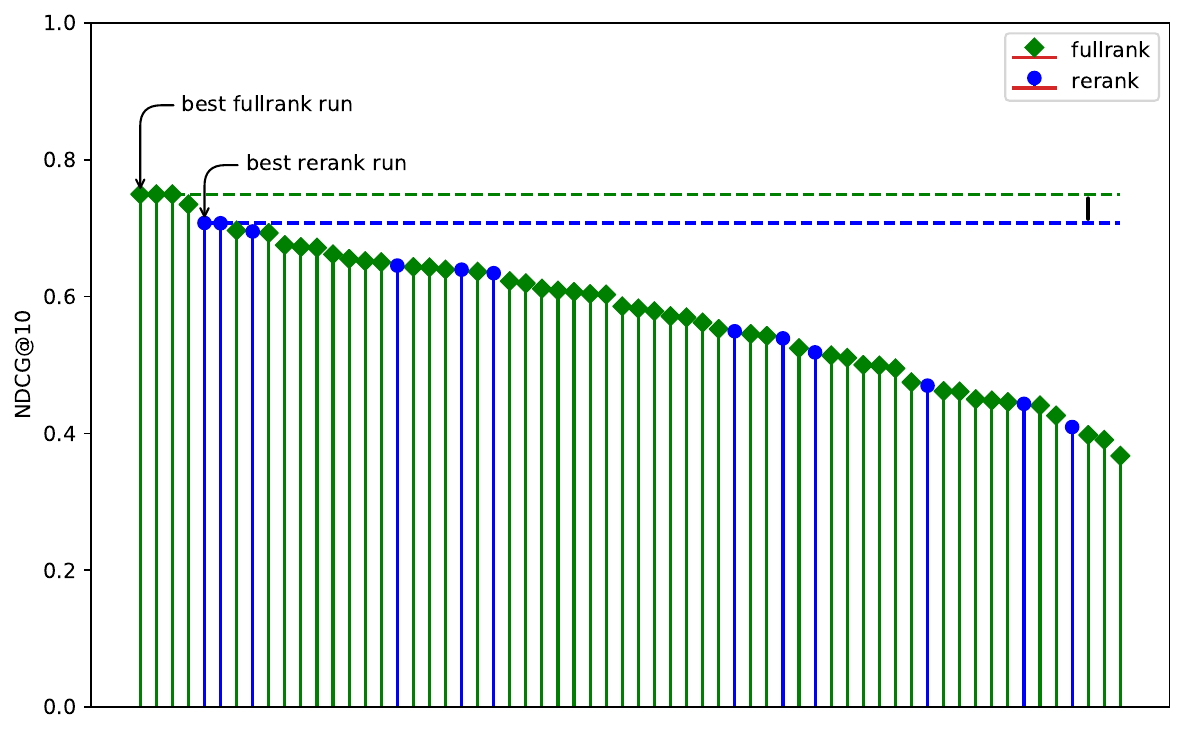}
    \caption{NDCG@10 for runs on the passage retrieval task}
    \label{fig:model-task-passages-stem-by-subtask}
  \end{subfigure}
  \begin{subfigure}{.49\textwidth}
    \includegraphics[width=\textwidth]{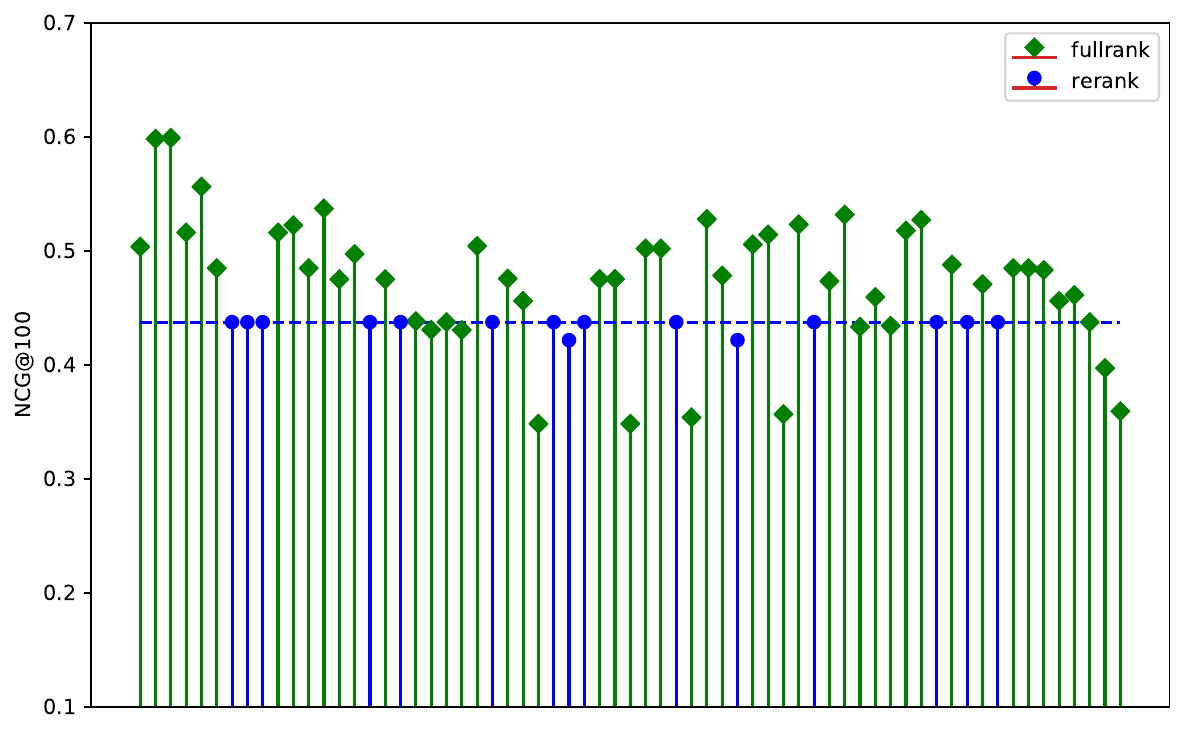}
    \caption{NCG@100 for runs on the document retrieval task}
    \label{fig:recall-task-docs-stem}
  \end{subfigure}
  \hfill
  \begin{subfigure}{.49\textwidth}
    \includegraphics[width=\textwidth]{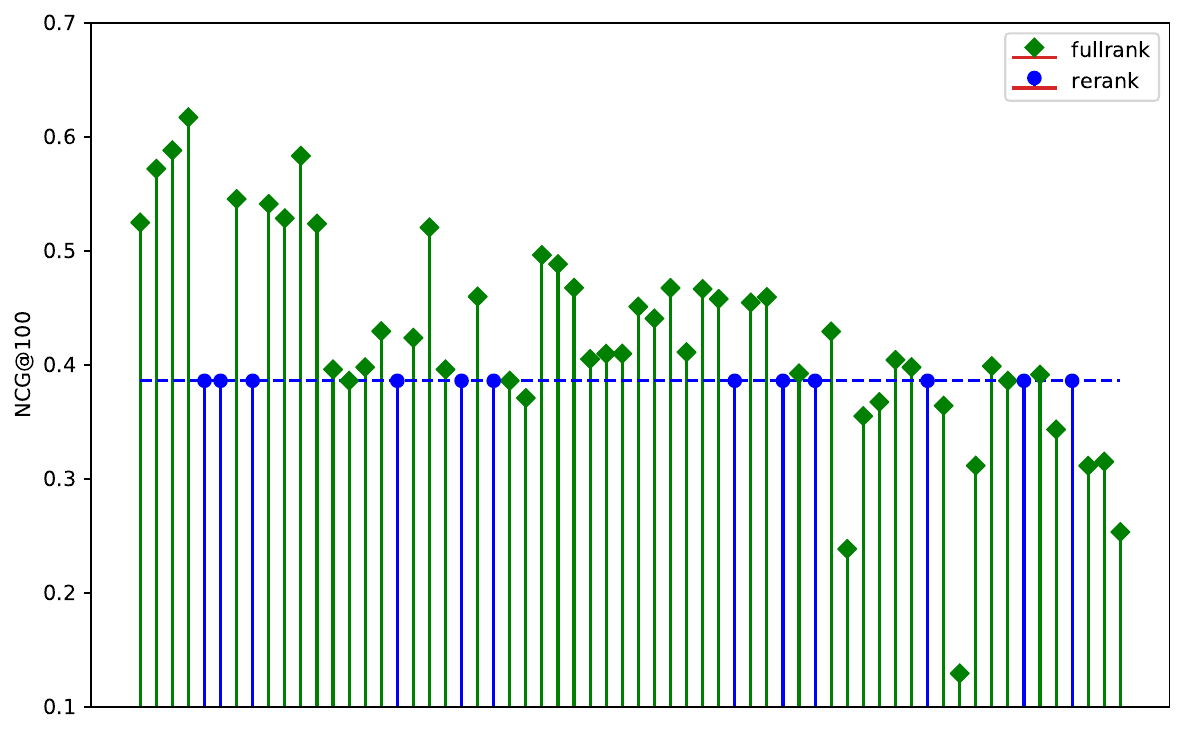}
    \caption{NCG@100 for runs on the passage retrieval task}
    \label{fig:recall-task-passages-stem}
  \end{subfigure}
  \caption{Comparing ``fullrank'' and ``rerank'' runs on ranking quality.
  Figure~(a) and (b) plots the NDCG@10 for different runs on the document and passage ranking tasks, respectively, and Figure~(c) and (d) plot the NCG@100 for the same.
  We order the runs by their NDCG@10 performance along the $x$-axis in all four plots.
  The best run for both tasks correspond to the ``fullrank'' setting.}
  \label{fig:recall-stem}
\end{figure}

\paragraph{End-to-end retrieval \vs reranking.}
This year for the document ranking task, the best ``fullrank'' run has a $4\%$ NDCG@10 improvement over the best ``rerank'' run, compared to $2\%$ in in 2019 and $5\%$ in 2020.
Similarly, for the passage task this year, the best ``fullrank'' run has $6\%$ higher NDCG@10 than than the best ``rerank'' run, which we can compare with a $4\%$ improvement in 2019 and no improvement in 2020.
If we compare Figure~\ref{fig:recall-stem} (b) and (d), we notice that a stronger correlation between NDCG@10 and NCG@100 metrics.
Also, this year the top-4 document ranking tasks and top-3 passage ranking tasks reported employing dense retrieval methods.
However, the jury is still out on whether neural methods have demonstrated gains under the full retrieval setting that leads to significant overall improvement in ranking quality.

This year we also analyze the performance gap between single stage retrieval methods and approaches that involve multiple stages of rank-and-prune~\citep{matveeva2006high, wang2011cascade}.
We find that single-stage retrieval methods do surprisingly well but still have a reasonable gap with the top run for both document ranking task ($6\%$ worse on NDCG@10) and the passage ranking task ($10\%$ worse on NDCG@10).

\begin{figure}
  \centering
    \includegraphics[width=.3\textwidth]{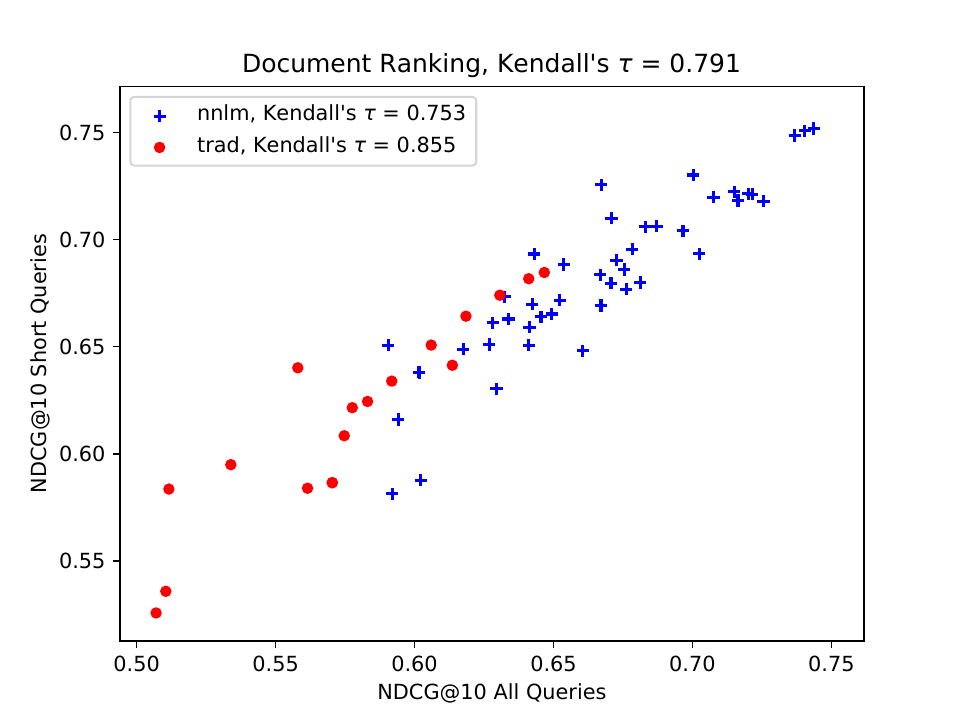}
    \includegraphics[width=.3\textwidth]{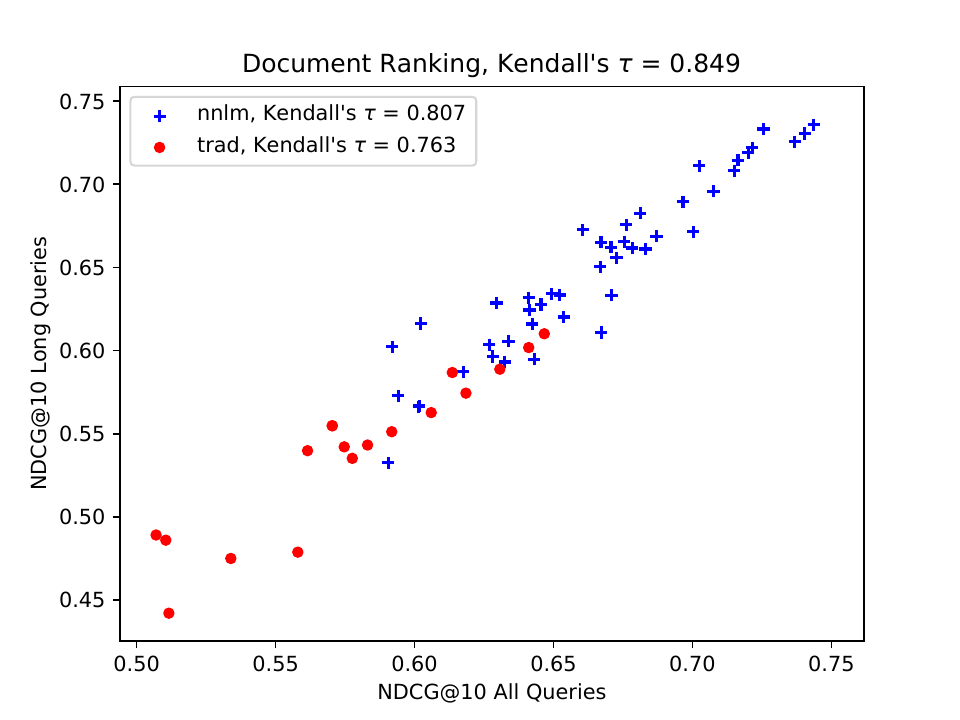}
    \includegraphics[width=.3\textwidth]{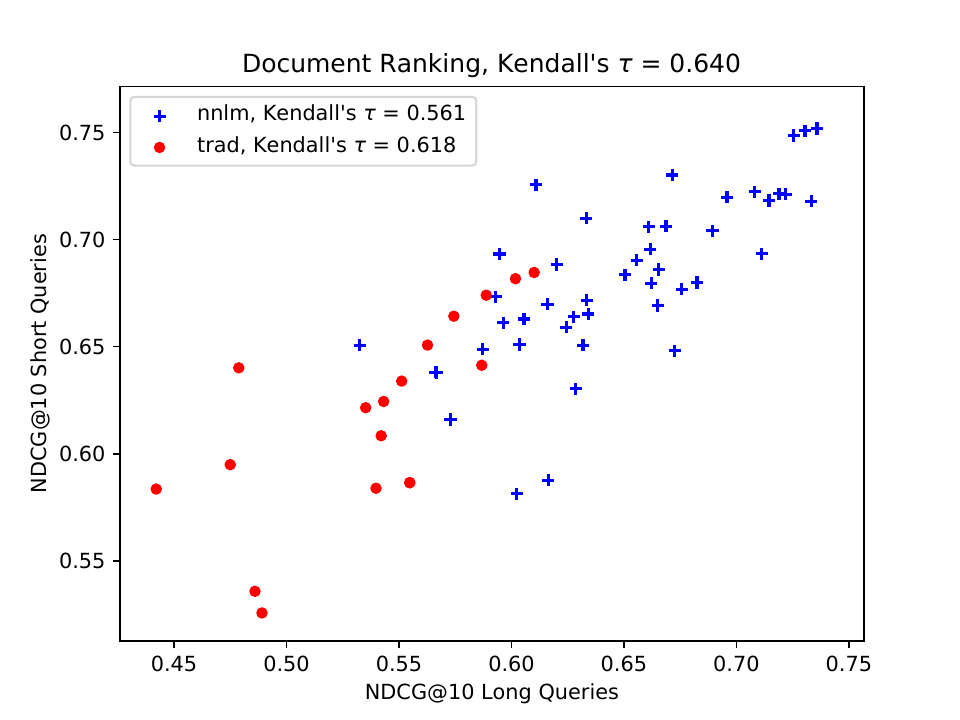} \\
    \includegraphics[width=.3\textwidth]{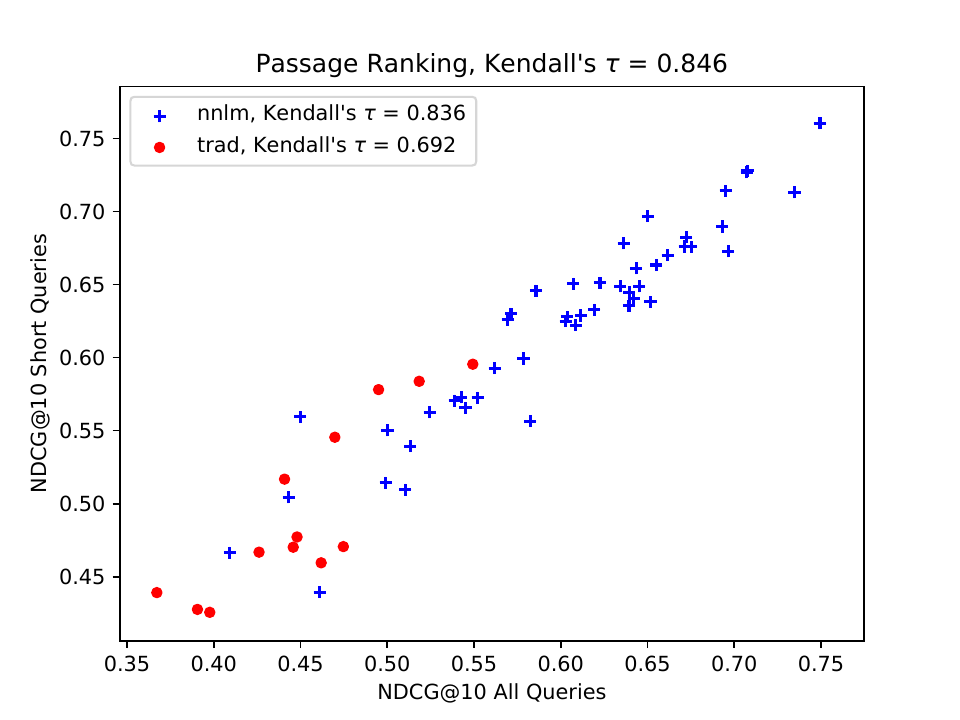}
    \includegraphics[width=.3\textwidth]{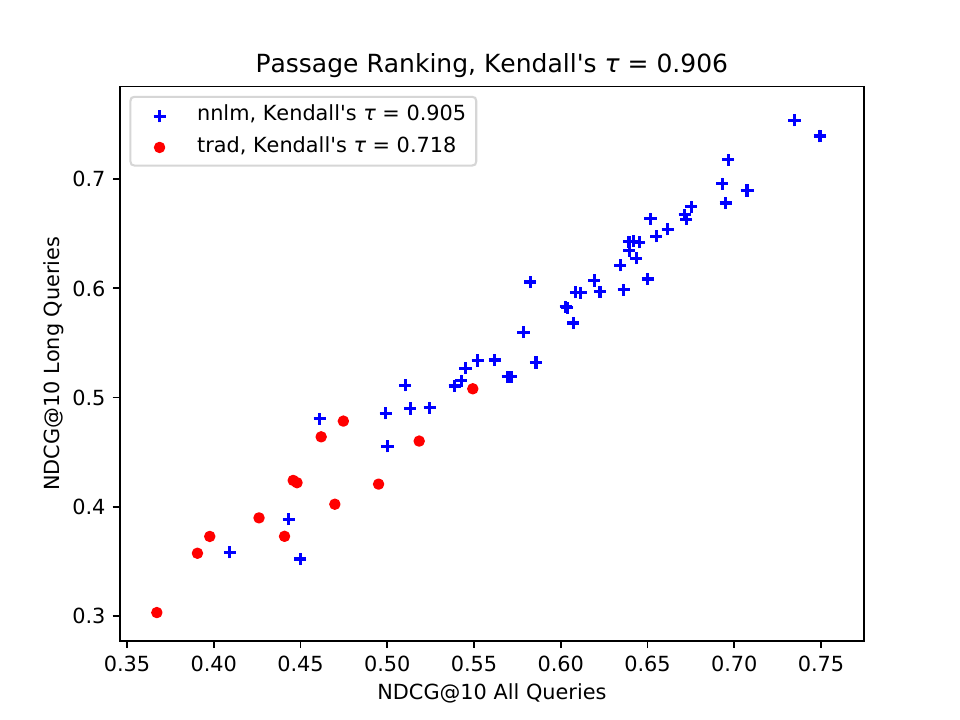}
    \includegraphics[width=.3\textwidth]{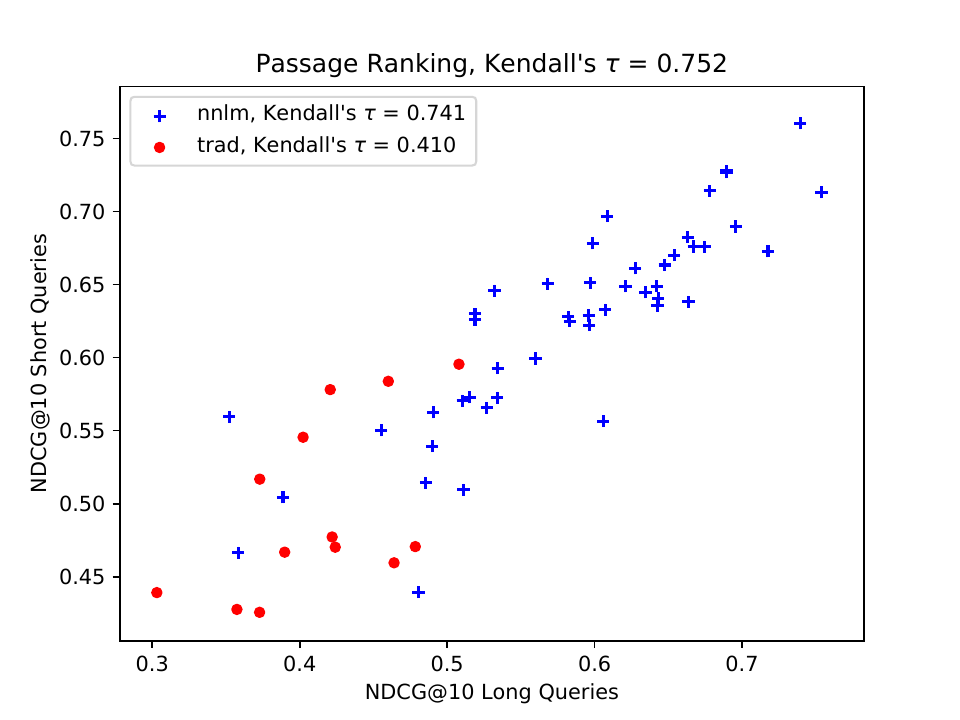}
    \caption{Kendall's $\tau$ correlation between ranking of systems using (left) short vs. all queries, (middle) long vs. all queries, and (right) long vs. short queries for (top) document retrieval and (bottom) passage retrieval tasks.}
    \label{fig:long-vs-short}
\end{figure}

\paragraph{Performance on long \vs short queries}
This year we used stratified sampling based on query length when selecting the queries for judging to analyze whether (1) the relative performance of ``nnlm'' and ``trad'' systems could change depending on query length, and 2) longer (hence, likely more difficult) queries are more discriminative in evaluating retrieval performance compared to shorter queries. Our test sets consisted of $57$ queries ($28$ short (number of words $< 10$), $29$ long (number of words $>= 10$) queries) for the document retrieval task and $53$ ($25$ short, $28$ long queries) queries for the passage retrieval task. 
We analyzed how the ranking of systems using long vs. short queries compare with ranking of systems using all the queries in the test set. Focusing on NDCG@10 as the evaluation metric, in Figure~\ref{fig:long-vs-short} we show the Kendall's $\tau$ correlation between the ranking of systems using (left) short vs. all queries, (middle) long vs. all queries, and (right) long vs. short queries for (top) document retrieval and (bottom) passage retrieval tasks. 

It can be seen that evaluation results obtained solely using long queries tend to be more correlated with results obtained using all the queries, suggesting that longer queries could be more discriminative than shorter queries for system evaluation. This behavior tends to be stronger for ``nnlm'' systems compared to ``trad'' systems. The rightmost plots in the figure show that correlations between ranking of systems using long vs. short queries tend to be relatively low, suggesting that relative performance of systems could vary significantly depending on the query length. Note that test sets for both document retrieval and passage retrieval tasks contain a slightly higher number of long queries compared to short queries, which could be affecting the conclusions reached. In the future, we plan to do further analysis controlling for the effect of the different number of queries. 

We further analyzed the number of relevant documents identified on average per query for long vs. short queries. Longer queries tend to be more specific; hence, we were expecting to identify fewer relevant documents for longer queries compared to shorter queries.  To our surprise, pools constructed for longer queries contained a higher number of relevant documents compared to shorter queries: $147.1$ vs. $140.6$ for document retrieval task and $67.8$ vs. $61.2$ for passage retrieval task. This could be another reason for the high correlation between evaluation results with long queries compared to all queries.

When absolute performance of ``trad'' systems across long vs. short queries are compared, it can be seen that ``trad'' systems tend to consistently perform worse on longer queries: the best performing trad run achieves $0.08$ higher NDCG@10 score on shorter queries compared to longer queries on the document retrieval task (NDCG@10 score of $0.69$ on shorter queries vs. $0.61$ on longer queries), and $0.09$ higher NDCG@10 on shorter queries for the passage retrieval task (NDCG@10 score of $0.60$ on shorter queries vs. $0.51$ on longer queries). On the other hand, the absolute performance of ``nnlm'' systems do not seem to be much affected from the query length. 

\paragraph{NIST labels \vs Sparse MS MARCO labels.}

\begin{figure}
    \centering
    \includegraphics[width=0.3\linewidth]{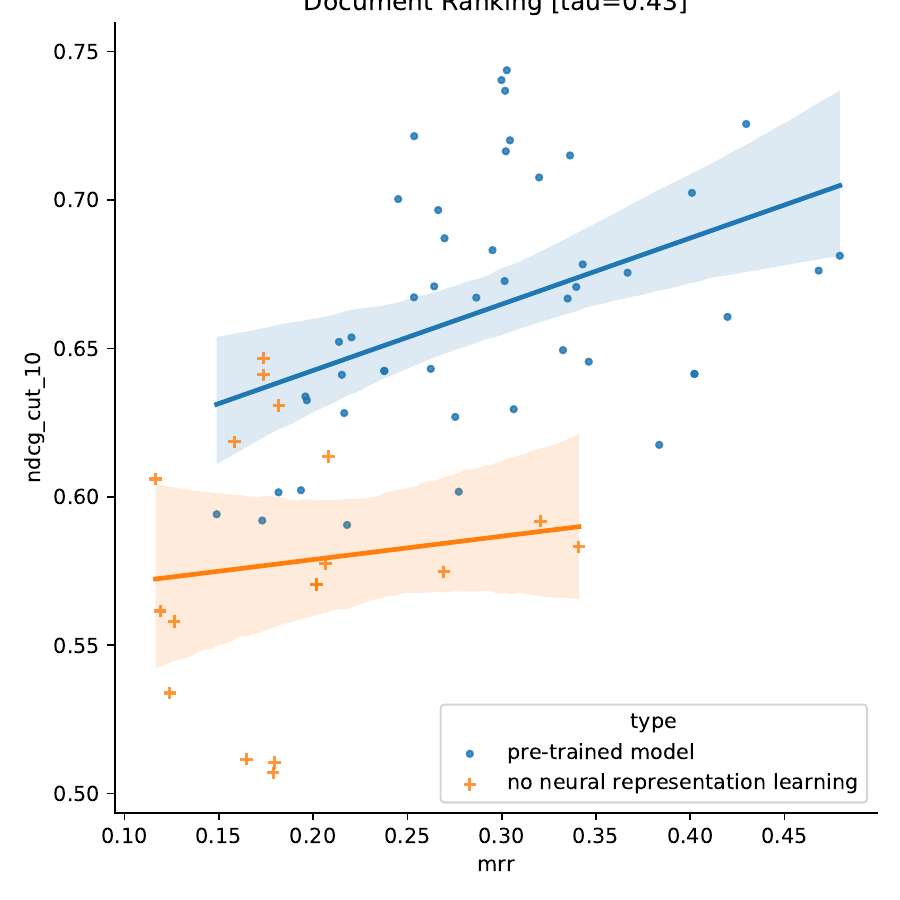}
    \includegraphics[width=0.3\linewidth]{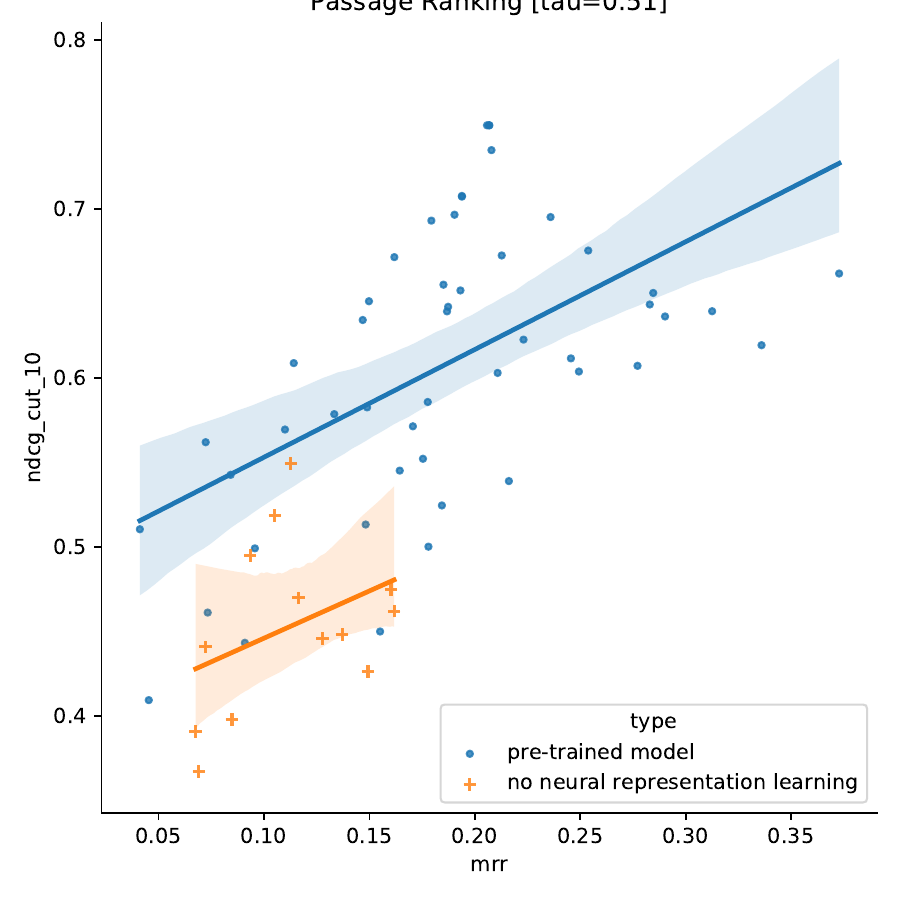} \caption{Agreement on evaluation outcomes when using sparse MS MARCO labels vs NIST labels.}
    \label{fig:rr_vs_ndcg}
\end{figure}

\begin{table}[]
\caption{NIST labels on positive MS MARCO qrels.}
\centering
\begin{tabular}{lllllllll}
\toprule
          & \multicolumn{4}{c}{NIST label} & \multicolumn{4}{c}{\% NIST label} \\
\cmidrule(r){2-5} \cmidrule(l){6-9}
          & 0      & 1     & 2     & 3     & 0      & 1      & 2      & 3      \\
\midrule
2020 docs & 11     & 12    & 7     & 16    & 24\%   & 26\%   & 15\%   & 35\%   \\
2020 pass & 7      & 12    & 15    & 23    & 12\%   & 21\%   & 26\%   & 40\%   \\
2021 docs & 4      & 11    & 11    & 35    & 7\%    & 18\%   & 18\%   & 57\%   \\
2021 pass & 7      & 12    & 13    & 14    & 15\%   & 26\%   & 28\%   & 30\%  \\
\bottomrule
\end{tabular}
\label{tab:nist_vs_marco_label_agreement}
\end{table}

The agreement between evaluation using sparse label RR (MS) and using full label NDCG@10 can be measured by comparing system ordering, for example calculating Kendall's Tau. Agreement is decreasing over the years of the track. In 2019, 2020 and 2021, the tau agreement on document ranking was 0.69, 0.46, and 0.43. On passage ranking 0.68, 0.69, and 0.51. This year's agreement plots can be seen in Figure~\ref{fig:rr_vs_ndcg}.

One reason for reduced agreement could have been that the dataset creation process for the v2 MS MARCO data applied positive labels to irrelevant results.
Unlike the document ranking dataset where we were able to transfer the v1 labels to the v2 collection using the document URLs as unique identifiers, for the passage dataset we had to transfer the labels based on matching the text content between v1 and v2 passages.
The latter raises potential for increased label noise that could negatively impact model training on the v2 dataset.
\citet{lassance2021naver} reported that when they performed a manual reassessment for a small sample of training queries, they find evidence of significant false positives and negatives in the v2 labels.
However, as a comparable analysis wasn't available for the v1 dataset, it is difficult to conclude how much of this may be explained by inter-annotator disagreement and general judgment noise.

Understanding the properties of the MS~MARCO v2 passage qrels could allow us to develop better training procedures, as measured using the official test data generated by NIST. Another way of analyzing the qrels, since the 2021 passage qrels are the ones in question, is to compare the qrels to NIST qrels, as in Table~\ref{tab:nist_vs_marco_label_agreement}. The table compares all cases where an evaluation query-result pair had both an MS~MARCO label and a NIST label. The analysis indicates that less than 20\% of this year's sparse labels were assigned label $0$ by NIST, which is in line with last year's numbers, we see no evidence from this that the positive labels have significantly decreased in quality in v2 data. For the document task, where the URL could be used to directly map the qrels, if anything the v2 data (2021) has better agreement between the two types of qrel. This could be because of the greatly improved quality of the document processing in the v2 data, giving documents with fewer character set issues, better extraction of main content and better layout. For the passage qrels, the 2021 data looks about the same or slightly worse, which could be understandable since the 2020 data is derived precisely from our human labeling task whereas 2021 data maps that label to a passage in the updated corpus.


\begin{figure}
    \centering
    \includegraphics[width=0.4\linewidth]{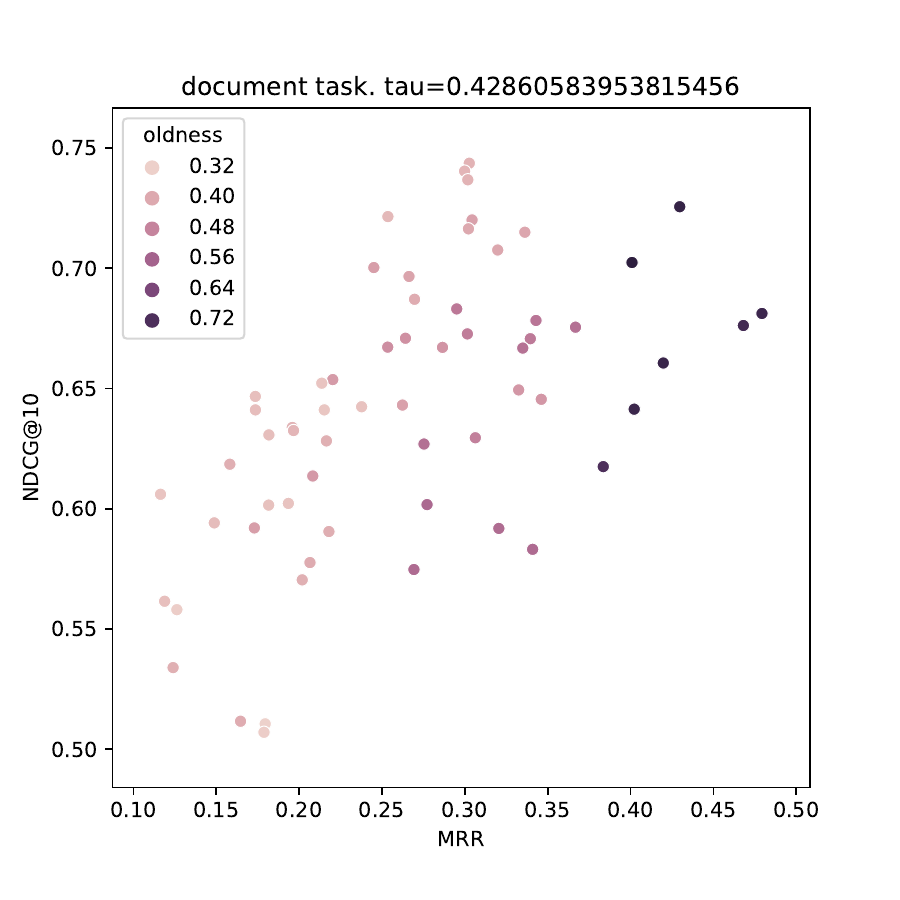}
    \includegraphics[width=0.4\linewidth]{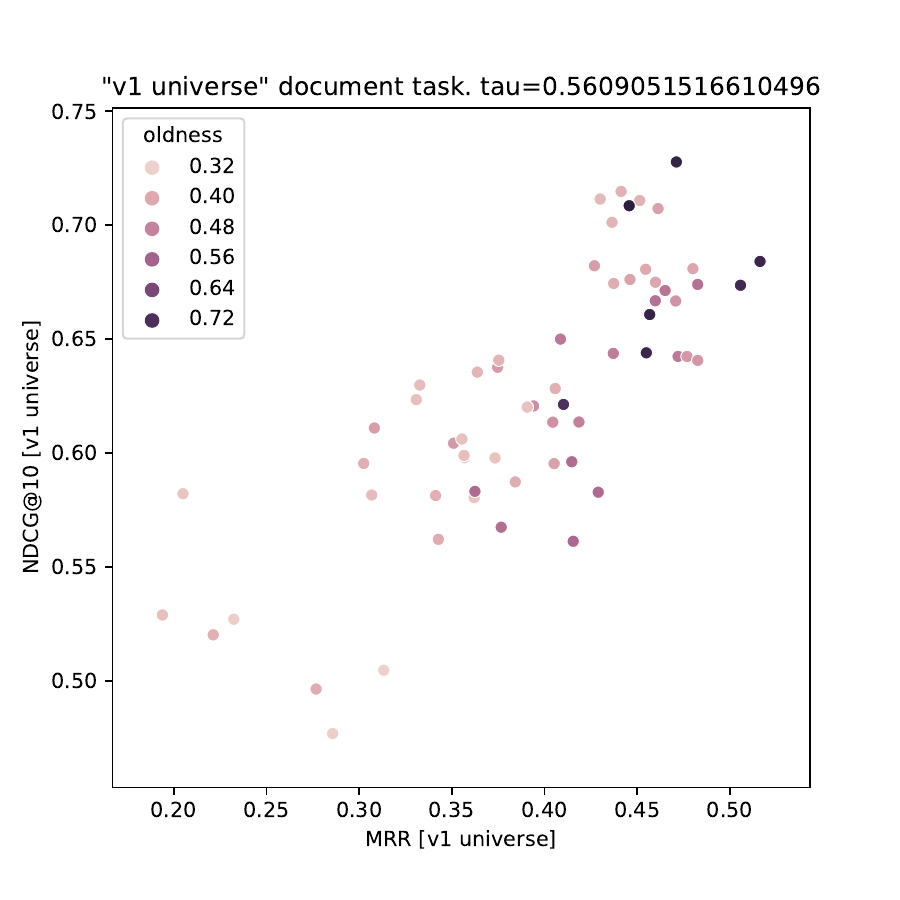}
    \caption{Metric disagreement based on `oldness'.}
    \label{fig:oldness}
\end{figure}

\paragraph{Oldness}

Another reason for reduced Tau agreement could be that some systems are learning something specific about the sparse MS MARCO data that's not true of NIST labels. We do find some evidence of this. For this analysis, we divide the document corpus into old and new URLs. The old URLs are the ones that are included in both v1 and v2 data. The new URLs are those that are present in v2 only. Note that both old and new URLs have document content from the present day, so if models can tell the difference, it's not because the old URLs have old content. Our hope was that old and new URLs would be indistinguishable, since they all use a snapshot of content from the present day, and that's important because in our MS MARCO sparse labels, only old URLs can have a positive label.

There are 2.7 million old URLs out of a total 12 million URLs in the v2 data. We define the metric oldness@10, such that a random ranker has an expected oldness@10 of $2.7/12=0.225$. However, because the v1 corpus was constructed based on the MS MARCO queries, and the new URLs were not selected based on those queries, we expect most runs to have higher oldness than a random run would. Because runs are paying attention to the query, and the old URLs have a higher density of relevant content. The question is whether some rankers learned, from training data, where all positive examples are old, to prefer old URLs. If so, their oldness@10 could be much higher than 0.225 Those systems with high oldness would also perform well on the MRR test data, since the sparse MRR test data also has the property that only old URLs can be positive. On the sparse MRR evaluation, returning a non-old document only has downside, since such documents can never be rewarded.

Figure~\ref{fig:oldness} shows the metric disagreement plot for the document task. It is like the plot in Figure~\ref{fig:rr_vs_ndcg}, but we colored each run according to its oldness. We can observe that most runs indeed have higher oldness than a random run. Some runs have very high oldness, of around 0.72. This suggests that the runs indeed learned to favor old documents. Those runs also perform particularly well on MRR, which only rewards old results, and this makes sense. These are top runs on MRR but not the best runs on NDCG@10. This is an artifact of the evaluation using sparse labels. We think that the runs with high NDCG@10 are the best runs.

To further dig into this, we create a version of every run that only uses old documents, and a version of the NIST qrels that only includes old documents. Of course, the MRR results are already old-only. We maintain the same coloring, to show which runs had high oldness before filtering. We call the after-filtering results the v1 universe, since it contains only the old documents and their associated qrels. The Tau agreement increases from 0.43 to 0.56, and the runs that had high oldness@10 are now in line with other runs.

This suggests multiple options for future use of the v2 data. One is to have some evaluation in the v1 universe. Another is that when running the dev set MRR, this should be done on the v1 universe. Another is that during training, the v1 universe should be used. This would mean that the positive training examples are unchanged, they are still all old. But when sampling negative documents during training, we no longer have an old-new pair, teaching the model to have high oldness. Instead, all the training pairs are now old-old, and the model purely learns about relevance ranking.

\paragraph{Completeness of judgments}

An artifact of the passage collection being more than 15 times larger and the document collection being more than three times larger in the v2 datasets is that the number of relevant passages and documents for the test queries also grew correspondingly.
As the NIST assessor budget has largely remained unchanged between the 2020 and the 2021 editions of the track, the increasing size of the relevance set lead to lower completeness of the NIST judgments.
This prompted the track organizers to issue a warning to the participants about high proportions of unjudged below the shallow pooling cutoff and potential implications for reuse of this dataset for benchmarking outside of the TREC settings.
One of the participating groups~\citep{kamps2021university} reported similar concerns as they noticed that the ratio of relevant-to-judged documents decreases much more gradually as we go down the ranks in the 2021 edition of the track compared to previous year.

\paragraph{Passage vs. document level judgments} 
\begin{table}[t]
\centering
\caption{Number of documents that have a particular inferred vs. actual label.}
\begin{tabular}{ll|cccc}
\multicolumn{6}{c}{\textbf{inferred rel}}\\
 & & 0 & 1 & 2 & 3\\ \hline 
\multirow{4}{*}{\rotatebox[origin=c]{90}{\textbf{actual rel}}} & 0 & 1096 & 186 & 49 & 12\\
 & 1 & 194 & 561 & 277 & 58\\
& 2 & 100 &	298	& 531 &	157 \\
& 3 & 49	& 99	& 190	& 425
\end{tabular}
\label{act_vs_inferred_qrels}
\end{table}

\begin{table}[t]
\centering
\caption{Percentage of documents with a particular inferred relevance grade, conditioned on the actual relevance label.}
\begin{tabular}{ll|cccc}
\multicolumn{6}{c}{\textbf{inferred rel}}\\
 & & 0 & 1 & 2 & 3\\ \hline 
\multirow{4}{*}{\rotatebox[origin=c]{90}{\textbf{actual rel}}} & 0 & $81.6\%$ & $13.8\%$ & $3.6\%$ & $0.9\%$\\ 
 & 1 & $17.8\%$ & $51.5\%$ & $25.4\%$ & $5.3\%$\\
& 2 & $9.2\%$ & $27.4\%$ & $48.9\%$ & $14.5\%$\\
& 3 & $6.4\%$ & $13.0\%$ & $24.9\%$ & $55.7\%$
\end{tabular}
\label{contingency_cond_actrel}
\end{table}

\begin{table}[t]
\centering
\caption{Percentage of documents with a particular actual relevance grade, conditioned on the inferred relevance label.}
\begin{tabular}{ll|cccc}
\multicolumn{6}{c}{\textbf{inferred rel}}\\
 & & 0 & 1 & 2 & 3\\ \hline 
\multirow{4}{*}{\rotatebox[origin=c]{90}{\textbf{actual rel}}} & 0 & $76.2\%$ & $16.3\%$ & $4.7\%$ & $1.8\%$\\ 
 & 1 & $13.5\%$ & $49.0\%$ & $26.5\%$ & $8.9\%$\\
& 2 & $6.9\%$ & $26.0\%$ & $50.7\%$ & $24.1\%$\\
& 3 & $3.4\%$ & $8.7\%$ & $18.1\%$ & $65.2\%$
\end{tabular}
\label{contingency_cond_infrel}
\end{table}

\begin{figure*}[t]
\centering
  \includegraphics[width=0.4\textwidth]{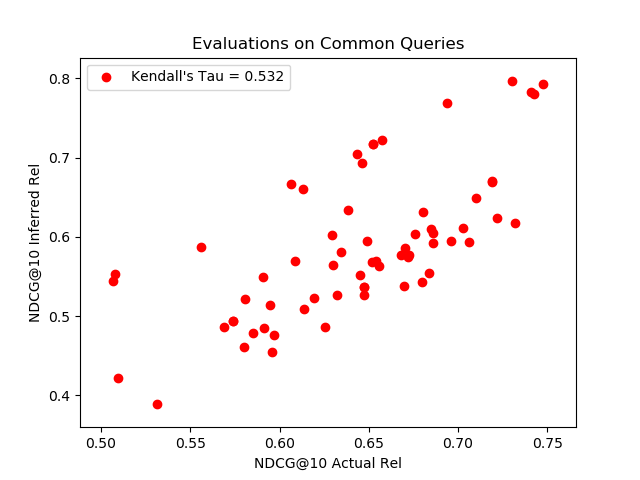}
  \includegraphics[width=0.4\textwidth]{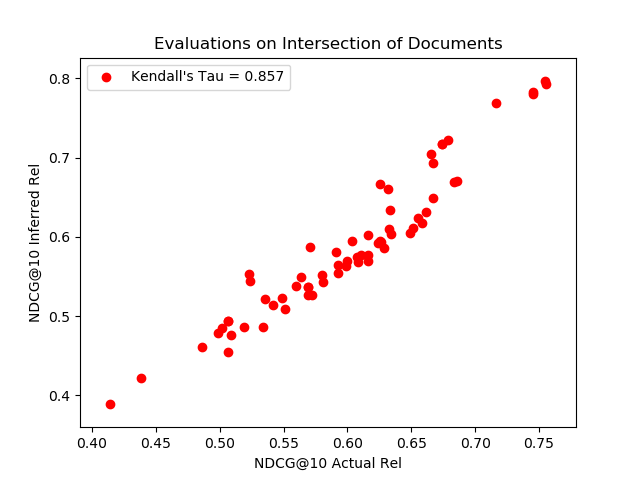}
  \vspace{-0.25cm}
  \caption{NDCG@10 values when systems are evaluated using actual labels vs. inferred labels when (left) all documents with labels are considered for evaluation for the common queries and (right) only documents that are common across the two judgments sets are used for evaluation.}
  \label{fig:evals}
  \vspace{-0.25cm}
\end{figure*}

During the construction of the passage ranking and document ranking test collections, the two collections were judged completely independently. Previous work showed that it may be possible to infer document level labels from passage level labels~\cite{Wu19} since passages that are relevant are likely to occur on relevant documents. If such a phenomenon holds for the Deep Learning Track test collections, it could be possible to mainly focus annotation effort on obtaining passage level labels and use them to infer document level labels. This could result in having a more complete test collection for the passage retrieval task, which could also result in a more completely judged document collection.

In order to analyze how reliable this process would be, we used the passage level labels to infer document level labels. Some documents contain several judged passages, and we used the maximum relevance score across all judged passages in a document to infer the final label for the document. During this process some documents did not have any inferred labels assigned to them as 1) the test collection for the passage retrieval task contains $7\%$ fewer queries than the document retrieval task, and 2) not all documents judged for the document retrieval task contain a passage that was judged for the passage retrieval task. 

Table~\ref{act_vs_inferred_qrels} shows how the document labels inferred from passage labels compare with the actual document labels, where each entry in the table shows number of documents with a specific inferred (columns) vs. actual (rows) relevance label. Figure~\ref{contingency_cond_actrel} shows the percentage of documents with a particular inferred relevance label, conditioned on the actual relevance label, and Figure~\ref{contingency_cond_infrel} shows the percentage of documents with a particular actual relevance label, conditioned on the inferred relevance label. It can be seen that while there is a reasonable agreement between the inferred labels and actual labels, there are also some differences between the two sets of labels. The inter-annotator agreement between the two sets of labels is $0.468$ according to Cohen’s Kappa, suggesting a moderate agreement between the two sets of labels.

We next compared how the evaluation results using inferred labels compare with evaluations using actual labels. Figure~\ref{fig:evals} shows the NDCG@10 values of systems submitted to the document retrieval task, with $x$ axis showing the metric value computed using the actual document labels compared to the metric value computed using the relevance labels inferred from passage level labels in the $y$ axis.  The left plot in the figure shows the evaluation results when all documents judged for the document retrieval task are used for computing the NDCG@10 value using actual judgments, ignoring the queries that were not included in the passage retrieval task. The right plot in the figure shows the results when the two evaluation results are computed on the intersection of the two judgments sets, ignoring the documents that do not have a corresponding inferred label. It can be seen that when the same documents are considered in evaluation, the two sets of judgments result in much higher agreement in terms of ranking of systems. This result suggests that while the document labels inferred from passage labels have the potential to be useful for evaluation, to compensate for the cases where no inferred label is available for a document (that should otherwise be annotated), we may need to obtain additional document level judgments. In the next year of the track, we might consider using such a hybrid evaluation dataset that contains a combination of inferred labels with actual labels in order to build a test collection that is more complete.


\section{Conclusion}
\label{sec:conclusion}

This is the third year of the TREC Deep Learning track.
This year we refreshed both the document and the passage collections consequently also growing both collections significantly in size.
We also continued to observe healthy participation in the track although the number of participating groups reduced slightly this year due to the delay in releasing the v2 collections.
Deep learning models with large scale pretraining continued to outperform traditional retrieval methods, and single stage retrieval with deep models seems to gain some more ground this year.
This report summarizes our analysis of submitted runs and the impact of the v2 collections.

\bibliographystyle{plainnat}
\bibliography{bibtex}

\end{document}